\documentclass{article}

\usepackage{microtype}
\usepackage{graphicx}
\usepackage{booktabs} % for professional tables

\usepackage{algpseudocode}
\usepackage{amsmath} % For mathematical symbols, if required

\usepackage{hyperref}
\usepackage{multirow}
\usepackage{subcaption}
\usepackage{float}
\usepackage{dblfloatfix}
\usepackage[ruled,vlined]{algorithm2e}

\usepackage[accepted]{icml2025}

\usepackage{amsmath}
\usepackage{amssymb}
\usepackage{mathtools}
\usepackage{amsthm}

\usepackage[capitalize,noabbrev]{cleveref}

\usepackage[textsize=tiny]{todonotes}

\icmltitlerunning{CIDD: Collaborative Intelligence Drug Design}

\begin{document}

\twocolumn[
% \icmltitle{Submission and Formatting Instructions for \\
%            International Conference on Machine Learning (ICML 2025)}
\icmltitle{Pushing the boundaries of Structure-Based Drug Design \\
through Collaboration with Large Language Models}

% \icmltitle{Streamlining Structure-Based Drug Design with Collaborative Intelligence}

% It is OKAY to include author information, even for blind
% submissions: the style file will automatically remove it for you
% unless you've provided the [accepted] option to the icml2025
% package.

% List of affiliations: The first argument should be a (short)
% identifier you will use later to specify author affiliations
% Academic affiliations should list Department, University, City, Region, Country
% Industry affiliations should list Company, City, Region, Country

% You can specify symbols, otherwise they are numbered in order.
% Ideally, you should not use this facility. Affiliations will be numbered
% in order of appearance and this is the preferred way.
\icmlsetsymbol{equal}{*}

% \begin{icmlauthorlist}
% \icmlauthor{Bowen Gao}{equal,yyy}
% \icmlauthor{Firstname2 Lastname2}{equal,air,thucs}
% \icmlauthor{Firstname3 Lastname3}{comp}
% \icmlauthor{Firstname4 Lastname4}{sch}
% \icmlauthor{Firstname5 Lastname5}{yyy}
% \icmlauthor{Firstname6 Lastname6}{sch,yyy,comp}
% \icmlauthor{Firstname7 Lastname7}{comp}
% %\icmlauthor{}{sch}
% \icmlauthor{Firstname8 Lastname8}{sch}
% \icmlauthor{Firstname8 Lastname8}{yyy,comp}
% %\icmlauthor{}{sch}
% %\icmlauthor{}{sch}
% \end{icmlauthorlist}

% \icmlaffiliation{air}{Department of Computer ,Tsinghua University}
% \icmlaffiliation{thucs}{Company Name, Location, Country}
% \icmlaffiliation{sch}{School of ZZZ, Institute of WWW, Location, Country}

% \icmlcorrespondingauthor{Firstname1 Lastname1}{first1.last1@xxx.edu}
% \icmlcorrespondingauthor{Firstname2 Lastname2}{first2.last2@www.uk}

\begin{icmlauthorlist}
\icmlauthor{Bowen Gao}{equal,aaa,bbb}
\icmlauthor{Yanwen Huang}{equal,ccc}
\icmlauthor{Yiqiao Liu}{ccc}
\icmlauthor{Wenxuan Xie}{ddd}
\icmlauthor{Wei-Ying Ma}{aaa}
\icmlauthor{Ya-Qin Zhang}{aaa}
%\icmlauthor{}{sch}
\icmlauthor{Yanyan Lan}{aaa}

\end{icmlauthorlist}

\icmlaffiliation{aaa}{Institute for AI Industry Research (AIR), Tsinghua University.}
\icmlaffiliation{bbb}{Department of Computer Science and Technology, Tsinghua University.}
\icmlaffiliation{ccc}{Department of Pharmaceutical Science, Peking University.}
\icmlaffiliation{ddd}{School of Future Technology, South China University of Technology.}

\icmlcorrespondingauthor{Yanyan Lan}{lanyanyan@air.tsinghua.edu.cn}

% You may provide any keywords that you
% find helpful for describing your paper; these are used to populate
% the "keywords" metadata in the PDF but will not be shown in the document
\icmlkeywords{Machine Learning, ICML}

\vskip 0.3in
]

% this must go after the closing bracket ] following \twocolumn[ ...

% This command actually creates the footnote in the first column
% listing the affiliations and the copyright notice.
% The command takes one argument, which is text to display at the start of the footnote.
% The \icmlEqualContribution command is standard text for equal contribution.
% Remove it (just {}) if you do not need this facility.

%\printAffiliationsAndNotice{}  % leave blank if no need to mention equal contribution
\printAffiliationsAndNotice{\icmlEqualContribution} % otherwise use the standard text.

\begin{abstract}
Structure-Based Drug Design (SBDD) has revolutionized drug discovery by enabling the rational design of molecules for specific protein targets. Despite significant advancements in improving docking scores, advanced 3D-SBDD generative models still face challenges in producing drug-like candidates that meet medicinal chemistry standards and pharmacokinetic requirements. These limitations arise from their inherent focus on molecular interactions, often neglecting critical aspects of drug-likeness. To address these shortcomings, we introduce the Collaborative Intelligence Drug Design (CIDD) framework, which combines the structural precision of 3D-SBDD models with the chemical reasoning capabilities of large language models (LLMs). CIDD begins by generating supporting molecules with 3D-SBDD models and then refines these molecules through LLM-supported modules to enhance drug-likeness and structural reasonability. When evaluated on the CrossDocked2020 dataset, CIDD achieved a remarkable success ratio of 37.94\%, significantly outperforming the previous state-of-the-art benchmark of 15.72\%. Although improving molecular interactions and drug-likeness is often seen as a trade-off, CIDD uniquely achieves a balanced improvement in both by leveraging the complementary strengths of different models, offering a robust and innovative pathway for designing therapeutically promising drug candidates.
\end{abstract}

\section{Introduction}
Structure-Based Drug Design (SBDD) is a keystone of modern rational drug discovery paradigm. Recently, various deep generative approaches have been applied in this field and have gained great advancements, enabling the direct generation of molecules for a given protein/pocket structure. Autoregressive models, such as AR~\citep{luo20213d} and Pocket2Mol~\citep{peng2022pocket2mol}, iteratively construct molecules by sequentially adding atoms to existing structures. While non-autoregressive generative models, including diffusion-based approaches like TargetDiff~\citep{guan20233d} and DecompDiff~\citep{guan2024decompdiff}, as well as Bayesian flow network-based models such as MolCRAFT~\citep{qu2024molcraft}, progressively decrease the noise from a provided random distribution to generate new molecules.

\begin{figure}[h]
\begin{center}
\centerline{\includegraphics[width=\columnwidth]{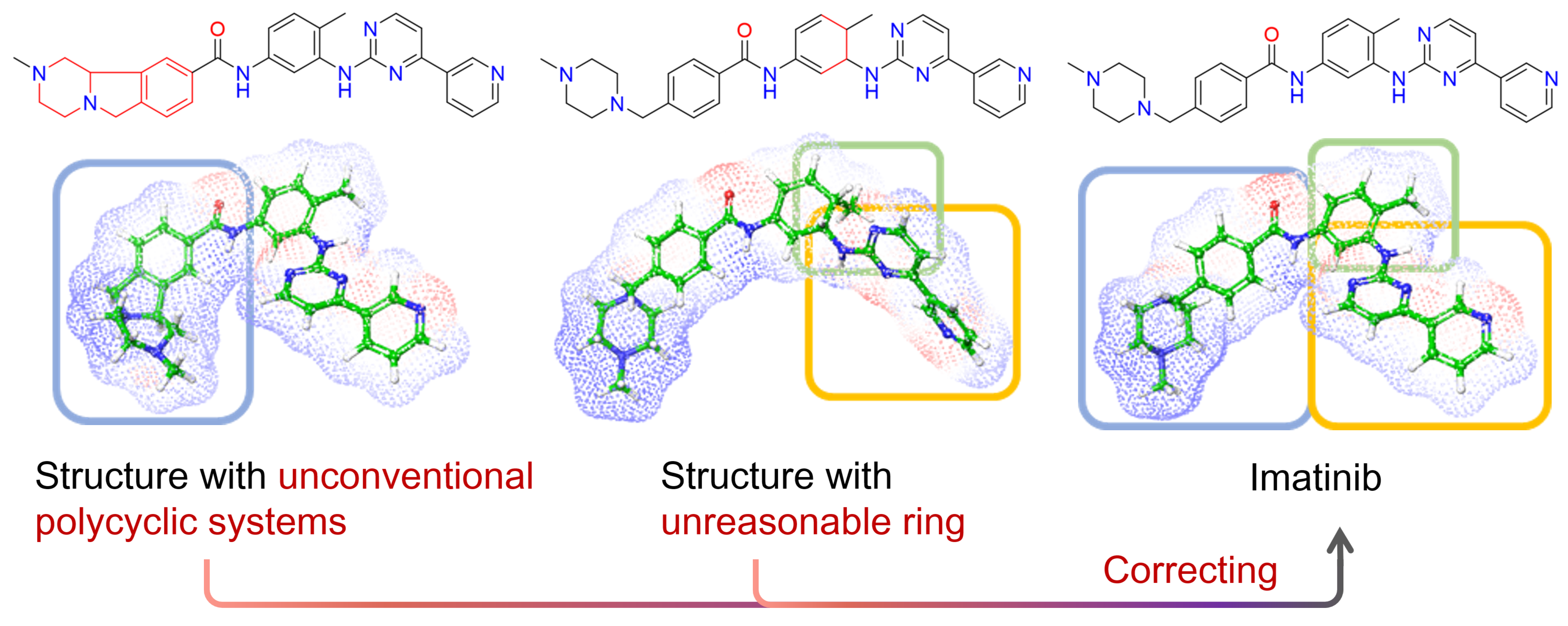}}
% \caption{\textbf{Two common issues in current SBDD model outputs}, seemingly minor 2D changes cause severe 3D structural disruptions. \textbf{(A)} The rationally designed anticancer drug Imatinib. \textbf{(B)} A common error where atoms are incorrectly connected into an unreasonable ring system. \textbf{(C)} Unreasonable conjugated arrangements in ring systems. }
\caption{\textbf{Two Common Errors in Current 3D-SBDD Model Outputs.} Distorted 3D substructures and their corresponding correct 3D conformations in the accurate structure are marked by boxes.}
\label{intro_case}
\end{center}
\vskip -0.2in
\end{figure}

Current 3D-SBDD models often achieve favorable docking scores by relying on distorted substructures, such as unconventional polycyclic systems or unreasonable ring formations, to fit target pockets. However, these distortions compromise molecular stability and reduce drug-likeness properties, such as aqueous solubility and oral absorption. As shown in Figure~\ref{intro_case}, introducing common SBDD errors into the rationally designed drug Imatinib results in substantial 3D conformational changes despite minimal 2D alterations. Correcting these distortions often disrupts the overall 3D structure, compromising binding affinity. This trade-off between structural accuracy and binding performance limits the practical utility of current 3D-SBDD models.

Addressing this issue within the existing SBDD framework is challenging due to the limitations of reconstruction objective. Specifically, current SBDD models focus on learning the distribution of $p(\text{molecule}\mid\text{target})$ from protein-ligand complex data, enabling the generation of molecules that exhibit rational structural bindings with the given targets. However, a significant gap remains between these molecules and their potential to be viable drugs, as they must also account for numerous complex factors, such as chemical reasonability, aqueous solubility, lipophilicity, binding affinity, pharmacokinetics, and more. These characteristics are clearly not easily captured through a conditional distribution $p(\text{molecule}\mid\text{target})$. Even if we can model them in a single distribution $p(\text{drug})$, which presents its own set of challenges, integrating both distributions into a unified framework for learning remains a complex task. As a result, solving this problem using SBDD models alone proves to be insufficient.

%, but they often overlook the characteristics of viable drugs encoded in distribution $p(\text{drug})$. As a result, 3D SBDD models inherently focus on adding atoms that can form interactions with the protein, without explicitly considering the plausibility and commonality of molecular fragments they generated. This results in molecules that may be valid in terms of spatial arrangements but lack the reasonability required for drug-likeness, as shown in Figure \ref{intro_case}.

%there is a significant gap between generated molecules and real-world drugs.

%deviate from the distribution of viable drugs $p(\text{drug})$, leading to a gap between designed molecules and real-world drugs. That is to say, those generative models are actually doing \textbf{structure-based molecule design} rather than \textbf{structure-based drug design}. Moreover, 3D SBDD models inherently focus on adding atoms that can form interactions with the protein, without explicitly considering the plausibility and commonality of molecular fragments they generated. This results in molecules that may be valid in terms of spatial arrangements but lack the reasonability required for drug-likeness, as shown in Figure \ref{intro_case}.

The recent emergence of large language models (LLMs) has led to significant breakthroughs across various scientific domains, including protein structure prediction \cite{valentini2023promises}, mathematical reasoning \cite{ahn2024large}, code generation \cite{hong2023metagpt}, and medical question answering \cite{lievin2024can}. These models leverage vast amounts of scientific literature, integrating a broad array of scientific knowledge, and have proven exceptionally effective in tackling complex scientific challenges. In drug discovery, for instance, models like GPT-4 have demonstrated impressive abilities in generating molecules with drug-likeness properties, achieving a 97.5\% ``reasonability" ratio compared to traditional SBDD models. Despite these promising results, LLMs face a critical limitation: they struggle to model the precise spatial arrangement of atomic coordinates in protein binding pockets. As a result, while these molecules may exhibit favorable drug-like characteristics, their binding affinities often fall short.

Inspired by these advances and the multifaceted challenges inherent in drug discovery, we propose a novel Collaborative Intelligence approach for drug design, termed CIDD. This approach combines the complementary strengths of 3D-SBDD models and LLMs to bridge the gap between binding affinity and molecular reasonability, ultimately enabling the generation of more effective drug candidates.
%molecules generated by the advanced LLM GPT-4o achieve an impressive reasonable ratio of 97.5\%. However, LLMs face a significant limitation: they struggle to effectively understand and model the spatial arrangement of atomic coordinates within protein binding pockets. Consequently, despite their drug-like characteristics, the generated molecules often exhibit poor binding affinities.

%Through this integration, LLMs are able to significantly enhance the reasoning capabilities required for solving intricate scientific problems.

%these models implicitly acquire valuable knowledge about molecular structures and properties essential for drug design. Our experiments highlight that molecules generated by the advanced LLM GPT-4o achieve an impressive reasonable ratio of 97.5\%. However, LLMs face a significant limitation: they struggle to effectively understand and model the spatial arrangement of atomic coordinates within protein binding pockets. Consequently, despite their drug-like characteristics, the generated molecules often exhibit poor binding affinities.

%The complementary strengths and weaknesses of 3D-SBDD models and LLMs inspire a novel framework that integrates their distinct intelligence. This framework, called Collaborative Intelligence Drug Design, combines the interaction-focused capabilities of 3D-SBDD models with the understanding of the property-structure relationship from LLMs to bridge the gap between molecular design and drug discovery. 

The process begins with 3D-SBDD models generating initial supporting molecules, which are refined through a streamlined pipeline consisting of several LLM-powered modules. An interaction analysis module identifies key molecular fragments contributing to crucial interactions with the protein pocket. A design module detects uncommon or suboptimal structures within the molecule and proposes modifications to enhance drug-likeness while preserving essential interactions. A reflection module evaluates prior designs, highlighting strengths and weaknesses to inform future designs. This cycle repeats multiple times to generate a variety of designs, which are then evaluated in a selection module to identify the optimal molecule balancing interaction capability and drug-likeness properties. By synergizing the structural interaction insights of SBDD with the extensive chemical expertise of LLMs, this framework enables the creation of molecules that excel in both target binding and human-preferred drug-like qualities.

We evaluate the performance of the CIDD framework on the CrossDocked2020 dataset \citep{francoeur2020three}, benchmarking it against several state-of-the-art (SOTA) 3D-SBDD models.
The results show that CIDD outperforms these models across multiple metrics, significantly improving both interaction capabilities and drug-likeness. Specifically, the CIDD framework increases the success ratio from 15.72\% to 37.94\%, and achieves an up to 16.3\% improvement in Docking Score, a 20.0\% boost in Synthetic Accessibility (SA) Score, an 85.2\% rise in Reasonable Ratio, and a 102.8\% increase in the ratio of molecules meeting multiple property requirements. These results highlight the ability of CIDD to optimize key molecular properties, particularly binding affinity and drug-likeness, underscoring its potential as a transformative tool in drug discovery.
% The Collaborative Intelligence Drug Design (CIDD) framework represents an innovative approach that integrates large language models as intelligent agents, seamlessly collaborating with human expertise, existing computational models, and advanced tools to drive pharmaceutical innovation. While this paper primarily focuses on refining molecules generated by SBDD models, the framework’s methodology extends to tackling critical challenges such as target identification, toxicity prediction during pre-clinical evaluation, and the design of molecular synthesis pathways. By combining human insight with the vast knowledge and coordination capabilities of LLMs, CIDD effectively leverages the strengths of machine learning models and tools. This transformative framework enhances every stage of the drug discovery process, paving the way for a fully automated and explainable drug development system that is both accessible and highly applicable for medicinal chemistry experts.
The CIDD framework integrates LLMs with human expertise, computational tools, and advanced models to accelerate pharmaceutical innovation. Focused on refining molecules from SBDD, CIDD can also be applied to target identification, toxicity prediction, and molecular synthesis, combining LLM knowledge with human judgment to optimize drug discovery and move toward an automated, explainable system for medicinal chemists.

\section{Preliminaries}

\subsection{Structure-Based Drug Design}

The goal of SBDD is to generate a molecule \(x\) that can bind to a given protein pocket \(P\). Recently, the development of SBDD models has shifted towards deep generative models. Autoregressive models, such as AR~\citep{luo20213d} and Pocket2Mol~\citep{peng2022pocket2mol}, generate molecules by sequentially adding new atoms to the existing structure. Subsequently, non-autoregressive generative models have emerged, including diffusion-based models like TargetDiff~\citep{guan20233d} and DecompDiff~\citep{guan2024decompdiff}, as well as Bayesian flow network-based models like MolCRAFT~\citep{qu2024molcraft}. 
%Additionally, fragment-based approaches such as FLAG have been introduced, which focus on assembling predefined molecular fragments to create new compounds.

\subsection{Large Language Models}

LLMs are advanced neural networks trained on vast corpora of textual data, enabling them to understand, generate, and process human-like language. Notable examples include GPT-4~\citep{achiam2023gpt}, LLaMA~\citep{touvron2023llama}, ChatGLM~\citep{glm2024chatglm}, and DeepSeek~\citep{liu2024deepseek}, which have demonstrated remarkable performance in tasks such as natural language understanding, code generation, mathematical problem solving, and logical reasoning. Their versatility and ability to learn complex patterns have made them increasingly relevant in domains beyond traditional natural language processing, including drug discovery~\citep{chakraborty2023artificial}.

Despite their potential, applying LLMs to structure-based drug design presents unique challenges. Protein structures are not purely textual but are spatially complex, involving three-dimensional arrangements of atoms and intricate chemical interactions. Representing these structural features in a form that LLMs can effectively process is a non-trivial task. 

\section{Methods}
\label{sec:methods}
% We first introduce how we design new metrics to evaluate whether a molecule is drug-like.
\subsection{New Metric Design}
\label{sec: new metrics}

% Drug discovery and development (DDD) is not purely rational and cannot be fully quantified. Nor can it be reduced to merely modifying a set of quantified targets. Instead, the DDD process is shaped by numerous human factors, including subjective preferences regarding molecular structures and varied task complexities, all of which draw on implicit or tacit knowledge that may be hidden in textbooks or the scientific literature. 
Drug discovery and development (DDD) is shaped by numerous human factors, including subjective preferences for molecular structures and the complexity of specific tasks. These factors are deeply rooted in implicit or tacit knowledge. 
% Consequently, even for experts in medicinal chemistry, it remains exceedingly difficult to fully characterize a function such as $p(drug)$.
However, this does not imply that human perceptions of ``drug-likeness" are indescribable or that $p(\text{drug})$ can simply be bent to match the output distributions of current molecular generative models. 

As highlighted in the preceding section, molecules generated by current SBDD models often diverge significantly from real drugs, particularly in the conjugation patterns of their ring systems. To capture this divergence, we developed two rule-based metrics—the Molecular Reasonability Ratio (MRR) and the Atom Unreasonability Ratio (AUR)—to assess ``drug-likeness", focusing on whether aromaticity is preserved in the examined molecules. Aromaticity, a fundamental concept in medicinal chemistry, describes the unique stability and electronic structure of certain ring systems, such as benzene. As a key feature of many Food and Drug Administration (FDA)-approved drugs, these structures are not only chemically stable but also essential for drug-target interactions, facilitating strong binding through mechanisms like $\pi$-$\pi$ stacking and hydrophobic interactions. However, current AI-driven generative models for SBDD often fail to replicate the nuanced use of aromatic rings observed in expert-designed molecules. These deviations lead to AI-generated molecules that significantly differ from clinically relevant drugs. By addressing this gap, MRR and AUR aim to better align AI-generated outputs with the practical and structural requirements of drug discovery.
% As highlighted in the preceding section, molecules generated by current SBDD models often diverge significantly from real drugs, particularly in the conjugation patterns of their ring systems. To capture this divergence, we developed two rule-based metrics, Molecular Reasonability Ratio (MRR), and Atom Unreasonability Ratio (AUR), to assess ``drug-likeness", focusing on whether aromaticity is preserved in the examined molecules. Aromaticity, a fundamental concept in medicinal chemistry, describes the unique stability and electronic structure of certain ring systems, such as benzene. As a key feature of many the Food and Drug Administration (FDA)-approved drugs, these structures are not only chemically stable but also essential for drug-target interactions, facilitating strong binding through mechanisms like $\pi$-$\pi$ stacking and hydrophobic interactions. However, current AI-driven generative models for SBDD often fail to replicate the nuanced use of aromatic rings observed in expert-designed molecules. These deviations lead to AI-generated molecules that significantly diverge from clinically relevant drugs. By addressing this gap, MRR and AUR aim to better align AI-generated outputs with the practical and structural requirements of drug discovery.

\paragraph{Molecular Reasonability Ratio.} The Molecular Reasonability metric evaluates the chemical plausibility of a molecule by analyzing its ring systems. Rings sharing one or more atoms are grouped into the same ring system, excluding carbonyl and imine groups. Each ring is evaluated to determine whether it forms an aromatic conjugated structure or a fully saturated ring. Rings meeting these criteria have their constituent atoms removed from further analysis. This process continues iteratively until no additional rings meet the criteria. If no ring atom(s) remain, the molecule is deemed reasonable; otherwise, it is considered unreasonable. Then for all molecules, we calculate the MRR. The full algorithm for MRR calculation is in Appendix \ref{sec: algorithm}.
% It flags molecules whose ring systems adhere to typical \emph{sp\textsuperscript{2}} and non-\emph{sp\textsuperscript{2}} arrangements, effectively reflecting how closely the structures match well-established hybridization norms;  
\paragraph{Atom Unreasonability Ratio.} The Atom-level Unreasonable is calculated as the number of atoms in the remaining rings that did not meet the criteria during the iterative process divided by the total number of atoms in all ring systems of the molecule under evaluation. This metric is averaged across all molecules, yielding the AUR.

These metrics provide practical and interpretable indicators of ``drug-likeness" from a structural perspective. By incorporating these metrics, AI-driven generative models can produce outputs that better align with real-world medicinal chemistry requirements. By focusing on key structural features such as ring systems, these metrics ensure that generated compounds not only meet computational criteria but also resonate with the empirical knowledge and intuitive preferences of human experts—a critical consideration for successful drug discovery.

\paragraph{QikProp Multiple Property Requirements.} To further evaluate the physicochemical and pharmacokinetic properties of the generated molecules, we employ QikProp, a tool recognized for its robust performance in predicting molecular drug-likeness properties~\citep{ioakimidis2008benchmarking}. The assessed properties include aqueous solubility, lipophilicity, polar surface area (PSA), the number of metabolizable sites, and oral absorption. Detailed requirements for each property are provided in Appendix \ref{sec:qkiprop}.

A molecule is considered to have passed the evaluation if it satisfies all $N$ predefined property requirements: $P_1, P_2, \dots, P_N$. If any of the properties fall outside the acceptable range, the molecule is classified as failing.
\[
\text{QikProp} = 
\begin{cases} 
1 & \text{if } P_1 \land P_2 \land \dots \land P_N \text{ are satisfied,} \\
0 & \text{otherwise.}
\end{cases}
\]
\begin{figure*}[ht]    
\centering
\includegraphics[width=1\textwidth]{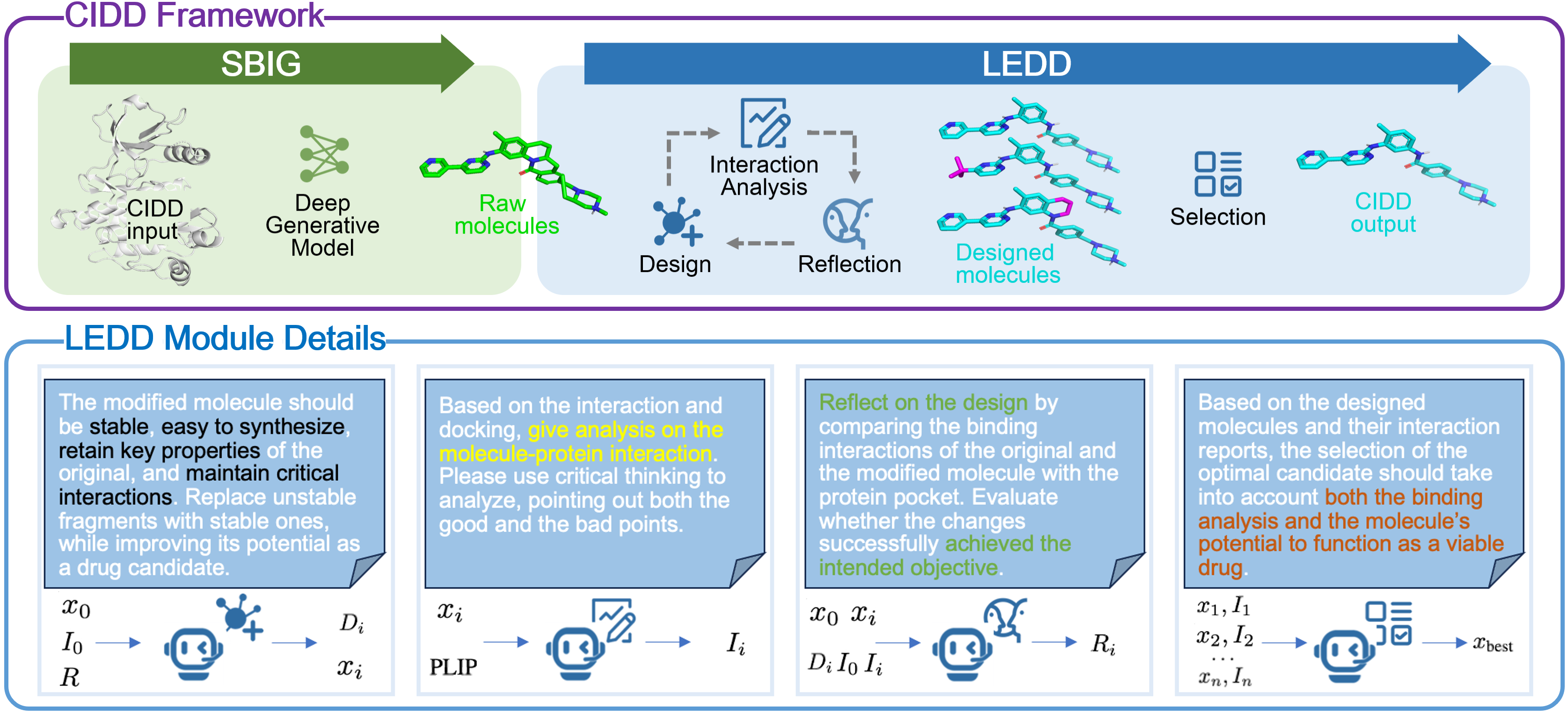}
\caption{\textbf{The CIDD Framework.} The framework comprises two main steps. First, the \textbf{Structure-Based Interaction Generator (SBIG)} generates a supporting molecule $x_0$ and its corresponding interaction report $I_0$. The second step, the \textbf{LLM-Enhanced Drug Designer (LEDD)}, refines the initial molecule. Using previous reflections $R = \{R_1, R_2, \dots, R_n\}$, the LEDD step produces a series of generated designs $D_i$, with corresponding designed molecules $X_i$ and interaction reports $I_i$. Finally, the selection module identifies the best-designed molecule, denoted as $x_{\text{best}}$. The prompts for each modules are also shown, with key words highlighted.}

\label{fig: main frame}
\end{figure*}

\subsection{CIDD Framework}

We propose the CIDD framework, illustrated in Figure~\ref{fig: main frame}, which integrates advanced computational techniques to advance the field of structure-based drug design. At its core lies the \textbf{Structure-Based Interaction Generator (SBIG)} step, which leverages 3D-SBDD models. These models excel at identifying and modeling interactions between protein pockets and small molecules, generating initial molecular proposals that adhere to the structural and chemical constraints of the target.

Following this initial step, the system employs a multi-module refinement process powered by LLMs, referred to as the \textbf{LLM-Enhanced Drug Designer (LEDD)}. This component is designed to transform the raw supporting molecular candidates into optimized drug-like compounds tailored to specific targets. By combining the detailed interaction capabilities of SBIG with the broad contextual knowledge of LEDD, the CIDD framework enables a transition from interaction-driven molecule design to comprehensive drug design.

The central objective of the LEDD component is to maintain the interaction capabilities of the original molecule while enhancing its drug-likeness. This transformation is modeled as a conditional distribution:
\[
x \sim P(\textit{Drug} \mid x_0, \textit{Target}),
\]
where \( x_0 \) represents the raw supporting molecule generated by SBIG, \(\textit{Target}\) is the protein pocket or biological target, and \( x \) is the refined drug-like compound. This probabilistic framework captures the refinement process, where initial molecular proposals are optimized into viable drug candidates designed for their specific targets.

A key challenge in this transformation lies in the lack of pairwise data linking unoptimized molecules to their drug-like counterparts. LLMs are particularly well-suited to address this challenge, as they are trained on extensive corpora of scientific literature and unstructured data. This training equips them with a nuanced understanding of chemical properties, preferred fragments, and drug-likeness criteria. By leveraging this knowledge, LLMs facilitate the generation of refined molecular structures that align with both target interaction requirements and drug development principles.

The integration of 3D-SBDD models and LLMs is a defining feature of the CIDD framework. While 3D models generate structurally sound initial proposals by focusing on interaction fidelity, LLMs improve these molecules by optimizing properties such as stability, solubility, and bioavailability. This complementary approach results in a coherent system that combines the strengths of both methodologies to advance the process of drug discovery.

The subsequent sections detail the complete pipeline of the CIDD framework and provide a thorough explanation of its modular components. This includes an examination of the interplay between SBIG and LLM-driven refinement, which underscores the potential of the CIDD system in addressing challenges in drug design.

% \begin{figure*}[ht]    
% \centering
% \includegraphics[width=1\textwidth]{fig/frame_0121.png}
% \caption{The Collaborative Intelligence for Drug Design(CIDD) framework.}
% \label{fig: main frame}
% \end{figure*}

\subsubsection{Pipeline}

The supporting molecules generated by the SBIG module are first processed by the \textbf{Interaction Analysis Module}, which produces a detailed interaction report outlining their binding characteristics with the target protein. This report, along with the initial supporting molecule, is then provided to the \textbf{Design Module}. The Design Module formulates improvement strategies based on the analysis of the supporting molecule and the interaction report, and applies these strategies to refine the molecule, resulting in a new, optimized version. The refined molecule is subsequently reanalyzed by the Interaction Analysis Module to assess its updated interaction properties.

The \textbf{Reflection Module} then takes the initial supporting molecule, the design steps, the modified molecule, and the interaction reports to evaluate whether the design has achieved its objectives. This process is repeated over $N$ rounds, with each round generating a new design proposal starting from the supporting molecule. Importantly, the reflections from previous rounds are used as guidance to inform and refine the design steps in subsequent rounds. 

Finally, the \textbf{Selection Module} evaluates all designed molecules and their corresponding interaction reports to select the best design that balances interaction quality and drug-likeness.

\subsubsection{Module Details}

\paragraph{Interaction Analysis.}

The \textbf{Interaction Analysis Module} evaluates the binding interactions between a molecule and the target protein pocket. Its input is a molecule, and its output is a detailed interaction report. The process is as follows:
1. Dock the molecule into the protein pocket.
2. Decompose the molecule into fragments using Breaking of Retrosynthetically Interesting Chemical Substructures (BRICS) \cite{degen2008art}.
3. Use Protein–Ligand Interaction Profiler (PLIP)~\citep{salentin2015plip} to identify all Non-Covalent Interactions between the molecule and the protein pocket.
4. Analyze the interaction results using an interaction-specific Large Language Model (\(LLM_I\)) to match each fragment to its interactions and assess their contributions. The process can be expressed as:
\[
    LLM_I\big(\text{PLIP}(\text{Docking}(x_i, P))\big) \quad \rightarrow \quad I_i
\]
Here, \(x_i\) is the molecule, \(P\) is the target protein pocket, and \(I_i\) is the interaction report generated by \(LLM_I\).

\paragraph{Design.}

The \textbf{Design Module} refines the supporting molecule \(x_0\) by proposing modifications based on its interaction report \(I_0\) and prior reflections \(R = \{R_1, R_2, \ldots, R_n\}\). A design-specific Large Language Model (\(LLM_D\)) formulates these design steps to enhance drug-likeness while maintaining the molecule’s binding efficacy. This process is expressed as:
\[
    LLM_D\big(x_0, I_0, R\big) \quad \rightarrow \quad D_i, x_i
\]
Here, \(D_i\) represents the design steps proposed by \(LLM_D\), and \(x_i\) is the modified molecule.

% \paragraph{Generation}

% The \textbf{Generation Module} implements the design steps proposed by the Design Module. A generation-specific Large Language Model (\(LLM_G\)) ensures that the resulting molecule is chemically valid and feasible. The process is formulated as:

% \[
%     LLM_G\big(D_i\big) \quad \rightarrow \quad x_i
% \]

% Here, \(x_i\) is the refined molecule generated by \(LLM_G\).

\paragraph{Reflection.}

The \textbf{Reflection Module} evaluates the design process by analyzing the initial molecule \(x_0\), the design \(D_i\), the refined molecule \(x_i\), and the interaction report \(I_0\) and \(I_i\). A reflection-specific Large Language Model (\(LLM_R\)) provides feedback for future iterations. This process is expressed as:
\[
    LLM_R\big(x_0, I_0, D_i, x_i, I_i\big) \quad \rightarrow \quad R_i
\]
Here, \(R_i\) is the feedback output generated by \(LLM_R\), used to improve subsequent design cycles.

\paragraph{Selection.}

The \textbf{Selection Module} evaluates all refined molecules \(x_1, x_2, \ldots, x_n\) and their interaction reports \(I_1, I_2, \ldots, I_n\). A selection-specific Large Language Model (\(LLM_S\)) determines the best candidate by balancing interaction quality and drug-likeness. The process can be expressed as:
\[
    LLM_S\big(\{x_1, I_1\}, \{x_2, I_2\}, \ldots, \{x_n, I_n\}\big) \quad \rightarrow \quad x_{\text{best}}
\]
Here, \(x_{\text{best}}\) is the optimal drug candidate selected by \(LLM_S\).

As shown in Figure \ref{fig: main frame}, we use domain-knowledge guided prompts to enable LLMs to rationally modify the molecules. All the detailed prompts and example responses for those modules are shown in Appendix \ref{sec:prompts}.

\section{Experiments}
\label{sec:experiment}

\subsection{Experiment Settings}

\paragraph{Dataset.}
Following the setup of previous 3D-SBDD models, we utilized the CrossDocked2020 dataset \citep{francoeur2020three} as our primary resource, and use the same splitting and testing setting as in TargetDiff \cite{guan20233d} that leads to 100 protein pockets for testing. 

\paragraph{Metrics.}
For evaluation, we adopted conventional metrics, including Vina docking score \citep{trott2010autodock} to assess the binding affinity of molecules, Quantitative Estimate of Drug-likeness (QED) score \citep{bickerton2012quantifying} to quantitatively evaluate drug-likeness, and SA score \citep{ertl2009estimation} to assess the synthetic feasibility of molecules. Additionally, molecular diversity was considered to measure the structural variability of the generated compounds, which is calculated by 1 - Extended-Connectivity Fingerprints 4 (ECFP4) similarity~\citep{rogers2010extended}.

Beyond these conventional metrics, we introduced a stronger focus on assessing the drug potential of molecules using novel metrics such as the MRR and AUR. These metrics provide deeper insights into molecular plausibility and atom-level chemical validity. Furthermore, we incorporated the QikProp pass ratio to refine the evaluation of drug-likeness, ensuring that the generated molecules meet critical physicochemical and pharmacokinetic criteria.

We also included the success ratio, as defined in previous studies, where success is determined by satisfying the criteria: Vina docking score \(< -8.18\), QED \(> 0.25\), and SA \(> 0.59\), as proposed by \citet{long2022zero}. Additionally, we use the proposed reasonability defined in section \ref{sec: new metrics} as an additional criterion for a molecule to be considered successful.
\paragraph{Baseline Models.}
We compared our CIDD method with several classic 3D-SBDD models, including the autoregressive models like AR and Pocket2Mol, diffusion-based models such as TargetDiff and DecompDiff, and the Bayesian flow network-based model MolCRAFT.
\paragraph{CIDD Settings.}
We utilize DecompDiff for the SBIG step. For the LEDD step, all modules are powered by GPT-4o. The Design Module operates for 5 rounds ($N=5$), generating 5 molecular candidates for the Selection Module, which then identifies the most optimal molecule. For each protein pockets, we generate 10 molecules. All the models used in SBIG step are trained with the CrossDocked2020 dataset and we use their released model weights.

\subsection{General Results}

The general benchmark results including CIDD and other 3D-SBDD models are shown in Table \ref{tab:main results}.

The presented CIDD utilizes the diffusion-based model DecompDiff to generate initial molecular proposals, which are then refined using our collaborative large language model system to enhance drug-likeness while maintaining interaction specificity. As shown in Table \ref{tab:main results}, CIDD outperforms conventional SBDD models, achieving an outstanding Vina docking score. It also surpasses all 3D-SBDD models in key druglikeness metrics, including MRR, AUR, and QikProp pass ratios. While CIDD slightly lags behind Pocket2Mol in QED and SA scores, its docking score far surpasses that of Pocket2Mol. This highlights the ability of CIDD to generate molecules that not only exhibit strong interaction potential but are also drug-like and synthetically feasible. CIDD significantly outperforms other methods in terms of success ratio, achieving \textbf{37.94\%} compared to \textbf{15.72\%} for the best alternative. This success ratio, which considers \textbf{docking score, synthetic feasibility, drug-likeness, and structural reasonability} together, highlights CIDD’s ability to overcome a key limitation of conventional \textbf{3D structure-based drug design (SBDD)} models—their difficulty in generating drug-like molecules. As a result, CIDD successfully facilitates the transition from \textbf{structure-based molecule design to structure-based drug design}, achieving our initial goal.

% \textbf{\begin{table*}[ht]
%     \setlength{\tabcolsep}{0.6em}
%     % \setlength{\tabcolsep}{6pt}
%     \vskip 0in
%     \caption{\textbf{Test Results on CrossDocked2020:} The best results are highlighted in \textbf{bold}, while the second-best results are indicated with \underline{underline}.}
%     \label{tab:main results}
%     \vskip 0.15in
%     \begin{center}
%     \scalebox{1.0}{
%     \begin{tabular}{l|lcccccccc}
%     \toprule
%     Method & Vina $\downarrow$ & QED $\uparrow$ & SA score $\uparrow$ & MRR $\uparrow$ & AUR $\downarrow$ & QikProp $\uparrow$ & Diversity $\uparrow$ & Success Ratio $\uparrow$ \\
%     \midrule
%     AR & -6.737 & 0.507 & 0.635 & 56.67\% & 34.72\% & 18.66\% & 0.836 & 3.28\% \\
%     Pocket2Mol & -7.246 & \textbf{0.573} & \textbf{0.758} & \underline{67.88\%} & \underline{20.14\%} & \underline{29.58\%} & 0.866 & 14.60\% \\
%     TargetDiff & -7.452 & 0.474 & 0.579 & 37.81\% & 43.40\% & 27.63\% & \textbf{0.890} & 3.04\% \\
%     DecompDiff & \underline{-8.260} & 0.444 & 0.609 & 62.60\% & 21.76\% & 29.04\% & \underline{0.877} & 15.72\% \\
%     MolCRAFT & -7.783 & 0.503 & 0.685 & 58.47\% & 25.59\% & 22.37\% & 0.870 & 13.72\% \\
%     \midrule
%     CIDD & \textbf{-9.019} & \underline{0.525} & \underline{0.694} & \textbf{76.54\%} & \textbf{11.44\%} & \textbf{37.54\%} & 0.870 & \textbf{37.94\%} \\
%     \bottomrule
%     \end{tabular}
%     }
%     \end{center}
%     \vskip -0.2in
% \end{table*}}

\begin{table*}[ht]
    \setlength{\tabcolsep}{0.6em}
    \vskip 0in
    \caption{\textbf{Test Results on CrossDocked2020:} The best results are highlighted in \textbf{bold}, while the second-best results are indicated with \underline{underline}.}
    \label{tab:main results}
    \vskip 0.15in
    \begin{center}
    \scalebox{1.0}{
    \begin{tabular}{l|lccccccccc}
    \toprule
    Method & Vina $\downarrow$ & QED $\uparrow$ & SA score $\uparrow$ & MRR $\uparrow$ & AUR $\downarrow$ & Success Ratio $\uparrow$ & QikProp $\uparrow$ & Diversity $\uparrow$ \\
    \midrule
    AR & -6.737 & 0.507 & 0.635 & 56.67\% & 34.72\% & 3.28\% & 18.66\% & 0.836 \\
    Pocket2Mol & -7.246 & \textbf{0.573} & \textbf{0.758} & \underline{67.88\%} & \underline{20.14\%} & 14.60\% & \underline{29.58\%} & 0.866 \\
    TargetDiff & -7.452 & 0.474 & 0.579 & 37.81\% & 43.40\% & 3.04\% & 27.63\% & \textbf{0.890} \\
    DecompDiff & \underline{-8.260} & 0.444 & 0.609 & 62.60\% & 21.76\% & \underline{15.72\%} & 29.04\% & \underline{0.877} \\
    MolCRAFT & -7.783 & 0.503 & 0.685 & 58.47\% & 25.59\% & 13.72\% & 22.37\% & 0.870 \\
    \midrule
    CIDD & \textbf{-9.019} & \underline{0.525} & \underline{0.694} & \textbf{76.54\%} & \textbf{11.44\%} & \textbf{37.94\%} & \textbf{37.54\%} & 0.870 \\
    \bottomrule
    \end{tabular}
    }
    \end{center}
    \vskip -0.2in
\end{table*}

\subsection{Improvments with Different Models}

\begin{figure}[h]
\begin{center}
\centerline{\includegraphics[width=1\columnwidth]{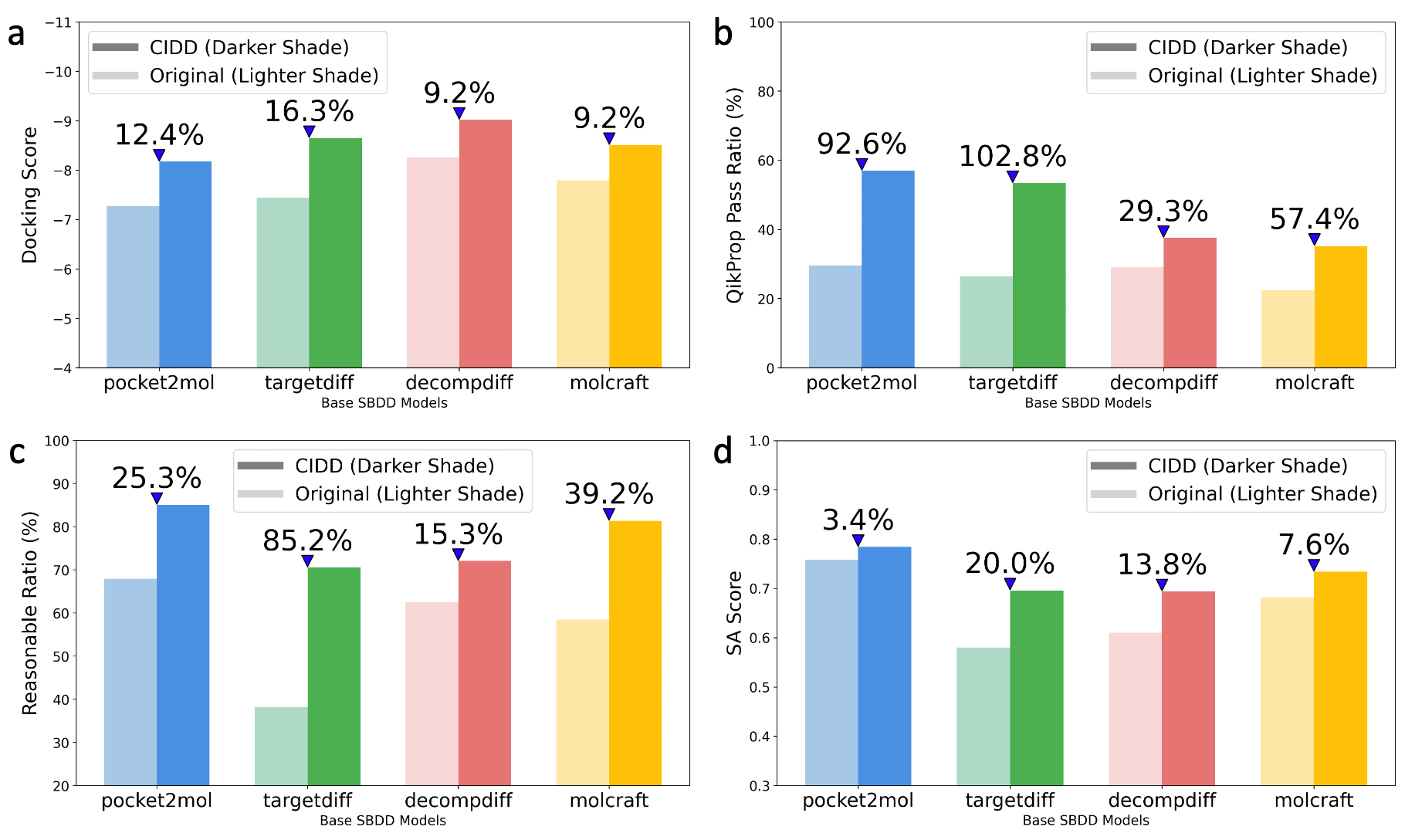}}
\caption{\textbf{Comparative Results:} (a) Docking Scores, (b) QikProp Ratios, (c) Molecule Reasonable Ratios, and (d) SA Scores before (\textit{lighter color}) and after (\textit{darker color}) applying the CIDD framework with different 3D-SBDD models.}
\label{fig: improvements}
\end{center}
\end{figure}

The proposed CIDD framework is a flexible and plug-and-play solution that enhances various 3D SBDD models by generating improved initial molecules. Figure \ref{fig: improvements} demonstrates the impact of CIDD across four key metrics: Docking Score (interaction quality), SA Score (synthetic accessibility), Reasonable Ratio (chemical fesibility), and QikProp Pass Ratio (general molecular properties). The lighter-shaded bars represent the baseline models, while the darker-shaded bars show performance after incorporating CIDD, with percentage improvements indicated between them. CIDD consistently enhances all metrics across models, achieving significant gains such as up to 16.3\% in Docking Score, 20.0\% in SA Score, 85.2\% in Reasonable Ratio, and 102.8\% in QikProp Pass Ratio. These results highlight the ability of CIDD to adapt and improve diverse SBDD models, particularly excelling in drug-likeness and molecular property optimizations, making it a robust and effective framework for drug discovery.

\subsection{Ablation and Analysis}
\subsubsection{Different LLMs}
For the CIDD results shown in Table \ref{tab:main results} and Figure \ref{fig: improvements}, the Large Language Model we used is \textbf{GPT-4o} from OpenAI. Here, we also test and compare different LLMs, including \textbf{GPT-4o-mini}, \textbf{GPT-4o}, and the recently released \textbf{DeepSeek-v3}~\citep{deepseekai2024deepseekv3technicalreport} and \textbf{DeepSeek-r1}\citep{guo2025deepseek}. The base SBDD model used in these tests is \textbf{MolCRAFT}.

The test results, presented in Table \ref{tab:llm ablation}, demonstrate that all tested LLMs effectively enhance the performance of the CIDD framework compared to using the base SBDD models alone. For metrics such as \textbf{MRR} and \textbf{QikProp}, which indicate drug-likeness, the LLMs show minimal differences. However, \textbf{GPT-4o}, \textbf{DeepSeek-v3} and \textbf{DeepSeek-r1} exhibit superior performance in improving docking scores, reflecting their enhanced ability to understand molecular interactions. Interestingly, the LLMs display notable variation in the similarity between the raw molecules proposed by the base SBDD models and the final modified molecules produced by CIDD. Since our goal is to minimize molecular modifications while improving properties, high similarity is preferred. This suggests that \textbf{GPT-4o-mini} may lack the ability to make subtle, targeted adjustments to molecular structures, whereas \textbf{DeepSeek-v3} emerges as the best performer, achieving property improvements with minimal modifications to the original molecules. Surprisingly, DeepSeek-r1 did not achieve better similarity metrics compared to v3, despite its superior performance in math problem-solving and complex reasoning tasks. We attribute this to the nature of our pipeline, which are more domain knowledge-driven rather than requiring complex reasoning. Furthermore, the results demonstrate that by leveraging models like GPT-4o and DeepSeek-v3, we can generate \textbf{a substantial amount of high-similarity pairwise data}, where one molecule exhibits improved properties over the other, as discussed in the previous section.
\vspace{-2pt}
\begin{table}[ht]
\centering
\caption{Ablation results for using different LLMs.}
\vspace{4pt}
\label{tab:llm ablation}
\resizebox{\linewidth}{!}{ % This will scale the table to fit the page width
\begin{tabular}{c|cccc}
\toprule
LLM & Vina$\downarrow$ & MRR$\uparrow$ & QikProp$\uparrow$ 
&Similarity$\uparrow$ \\ 
\noalign{\vskip 4pt}
\hline
\noalign{\vskip 4pt}
-&  -7.78   & 58.47\%    & 22.37\%  & - \\
\hline
\noalign{\vskip 4pt}

GPT-4o-mini &  -8.29   & 80.02\%    & 36.43\%  & 0.220 \\
GPT-4o &  -8.50    & 81.37\%  &   35.22\%   & 0.296 \\
DeepSeek-v3  & -8.49   & 76.00\%  & 34.13\%  & 0.379 \\
DeepSeek-r1  & -8.57   & 79.17\%  & 36.72\%  & 0.182 \\
\noalign{\vskip 2pt}
\bottomrule
\end{tabular}
}
\end{table}

\subsubsection{Number of Designs}

\begin{figure}[h]
\begin{center}
\centerline{\includegraphics[width=0.8\columnwidth]{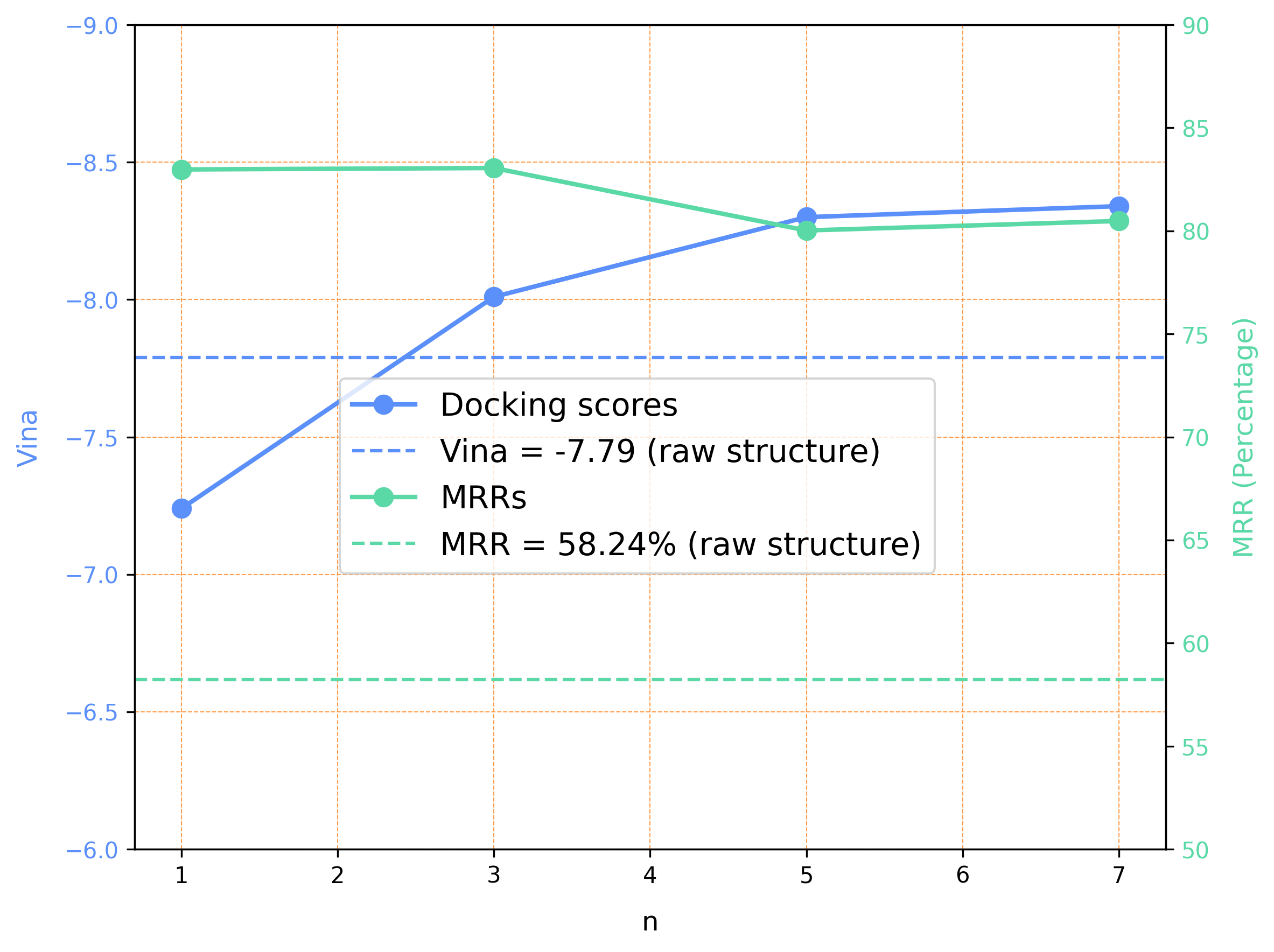}}
\caption{\textbf{Ablation Study on the Number of Designs:} The x-axis represents \( n \), the number of designs. The \textcolor{blue}{blue line} corresponds to \textbf{Vina docking scores}, while the \textcolor{green}{green line} represents \textbf{Molecule Reasonable Ratios}. \textbf{Dashed lines} indicate the values for the \textbf{initial supporting molecule} generated by the 3D-SBDD model.}
\label{fig: ablation designs}
\end{center}
\vspace{-5pt}
\end{figure}

An important hyperparameter in CIDD is the number of proposed designs (\(N\)). To evaluate its impact, we conduct experiments with varying values of \(N\). As illustrated in Figure~\ref{fig: ablation designs}, the x-axis represents the number of designs, while the y-axis displays the Vina docking score and the MRR. Dashed lines indicate the corresponding values for the original supporting molecules. When \(N=1\), the average docking score is slightly worse than that of the supporting molecules. However, the molecular reasonability ratio already shows a significant improvement. As \(N\) increases, the docking score improves progressively, although the MRR experiences a slight decline. At \(N=5\), a favorable balance is achieved, with both the average docking score and MRR better than the original supporting molecules.
\begin{figure*}[h]
\begin{center}
% \centerline{\includegraphics[width=1\linewidth]{fig/cidd_case_tmp.png}}
\centerline{\includegraphics[width=\linewidth]{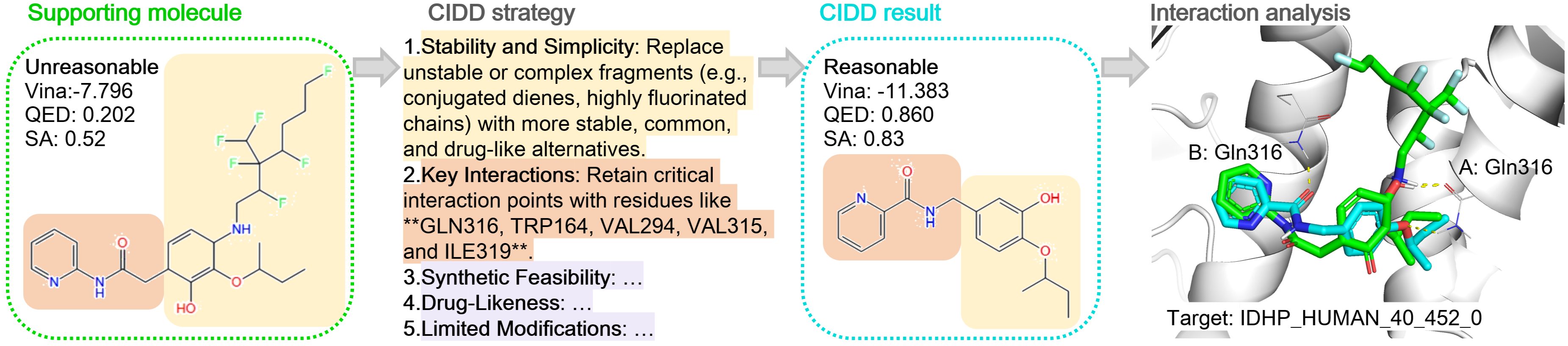}}
\caption{\textbf{Generation Case and Corresponding Design Strategy Produced by CIDD.}}
\label{fig: designs}
\end{center}
\end{figure*}
\subsubsection{Using Pure LLM for Generations}
\begin{table}[ht]
\centering
\caption{Ablation study of using pure LLM for SBDD.}
\vspace{4pt}
\label{tab:llm only ablation}
\resizebox{1\linewidth}{!}{ % This will scale the table to fit the page width
\begin{tabular}{c|cccc}
\toprule
 & Vina$\downarrow$ & MRR$\uparrow$  &Diversity$\uparrow$ &Success Ratio$\uparrow$ \\ 
\noalign{\vskip 4pt}
\hline
\noalign{\vskip 4pt}
LLM-SBDD &  -6.244   & 97.45\%    & 0.808 &5.95\% \\
\noalign{\vskip 4pt}
CIDD-LLM&  -7.230    & 90.97\%  &  0.830 & 17.59\%   \\
\hline
\noalign{\vskip 4pt}
CIDD & -9.019   & 76.54\%  & 0.870 & 37.94\%  \\

\noalign{\vskip 2pt}
\bottomrule
\end{tabular}
}
\end{table}
As previously discussed, using pure LLMs for structure-based drug design is challenging due to their inability to comprehend the complex three-dimensional structure of protein pockets. To validate this, we conducted a test by providing the pocket structure in PDB format to the LLM and instructing it to generate molecules capable of binding to the target. This approach is referred to as LLM-SBDD. Additionally, we integrated the LLM-SBDD into the CIDD framework, using the LLM-SBDD for the SBIG step. This variant is referred to as CIDD-LLM.

As shown in Table~\ref{tab:llm only ablation}, LLM-SBDD generates molecules with a very high reasonability ratio, supporting our motivation for using LLMs to refine supporting molecules into drug candidates. However, LLM-SBDD produces molecules with poor Vina docking scores, resulting in a low overall success ratio compared to the standard CIDD, which utilizes DecompDiff as the SBIG step. These results highlight the critical importance of collaborative intelligence, leveraging both the 3D-SBDD models for their ability to construct interactions and LLMs for their extensive chemical knowledge.

\subsection{Analysis of Generated Molecules}
Figure~\ref{fig: designs} illustrates the CIDD generation process. The LLM-powered modules within the CIDD framework analyzed and refined the raw supporting molecule (green), resulting in a high-quality final structure (blue). The modules automatically detected issues in the supporting molecule, such as an unreasonable diene substructure, which was replaced with a benzene ring, and an uncommon fluorinated chain, which was modified accordingly. Additionally, side chains were optimized to maintain hydrogen bonding with Gln316 on both Chain A and B of the target protein, improving the original docking score as planned while enhancing the overall drug-likeness properties. This also demonstrates another advantage of our method: \textbf{the design process is explainable.} CIDD not only generates optimized molecules but also provides insights into the rationale behind its design, highlighting structure strengths and potential areas for improvement. This transforms traditional opaque SBDD into an explainable and interpretable approach, making it more beneficial for human experts in drug discovery.

% Figure~\ref{fig: designs} shows a case that illustrates the generation process of CIDD. As shown in the figure, the LLM-powered modules within the CIDD framework performed a rational analysis and devised strategies to refine the supporting molecule, producing a high-quality final structure. In this example, the LLM-powered module automatically detected the issues within the supporting molucule and generated the strategy in the figure. It successfully identified that the diene substracture is unreasonable and should be changed to beneze as did in the resulting structure, and it also identified that the fluorinated chain is uncommon and thus changed it accordingly. Furthermore, the side chains were strategically optimized to meet the strategy it developed that the hydrogen bond interaction with the Gln316 on both Chain A and B of the target protein been retained to largely improve the original vina score when enhancing the overall drug-like properties of the molecule. This also demonstrates another advantage of our method: \textbf{the design process is explainable.} CIDD not only generates optimized molecules but also provides insights into the rationale behind its design, highlighting strengths and potential areas for improvement. This transforms traditional black-box SBDD into an explainable and interpretable approach, making it more beneficial for human experts in drug discovery.

As demonstrated by the visualizations and the prior ablation study, there is a high degree of similarity between the initial raw molecule and the fine-grained molecule generated by CIDD. These modifications are controllable, as they primarily involve substituting unreasonable fragments with reasonable counterparts, rather than transforming the entire molecule into something entirely different. This provides significant advantages from a \textbf{data-centric perspective.} This approach enables the \textbf{automatic generation of extensive pairwise data}, where one molecule is a refined version of another, exhibiting enhanced drug-likeness properties. Such data inherently captures the transformation of the distribution \(P(\text{Drug} \mid \text{molecule}, \text{target})\), effectively addressing the data scarcity challenge identified earlier. More generation cases for CIDD are shown in Appendix \ref{sec: more cases}.

Utilizing this pairwise data enables fine-tuning of existing 3D-SBDD models, enhancing their ability to generate molecules that are both chemically plausible and meet multi-objective criteria for valid drugs. Unlike naive methods that modify a ``good" molecule to create a ``bad" version with a random distribution, our approach ensures that the distribution of ``bad" molecules aligns closely with the output distribution of 3D-SBDD models. This alignment makes the data more compatible with existing frameworks, representing a significant contribution from a data-centric perspective.

% \section{Conclusion}
% In this paper, we propose the CIDD framework, which addresses key limitations of traditional SBDD by integrating advanced 3D-SBDD models, adept at molecular interactions, with LLMs, skilled in understanding drug requirements. Experimental evaluations on the CrossDocked2020 dataset demonstrate the superior performance of CIDD, significantly improving the success ratio of generated molecules considering both interaction ability and drug-likeness. compared to state-of-the-art models.This work pioneers the collaboration between domain-specific models and LLMs, advancing toward a more automated, explainable, and effective drug discovery pipeline. While in this work the primary focus is molecule generation, such idea and framework can be further extended to other steps of the drug discovery pipeline, like target identification, hit to lead optimization, pre-clinical toxity evaluation and synthetic pathway planning. We believe the collaboration between task-specific machine learning models and Large Language Models can lead to a better future and more promising, effective and explainable drug discovery pipeline.

\section{Research Rationale and Contribution Highlights}

\begin{figure*}[ht]
\begin{center}
\centerline{\includegraphics[width=0.8\textwidth]{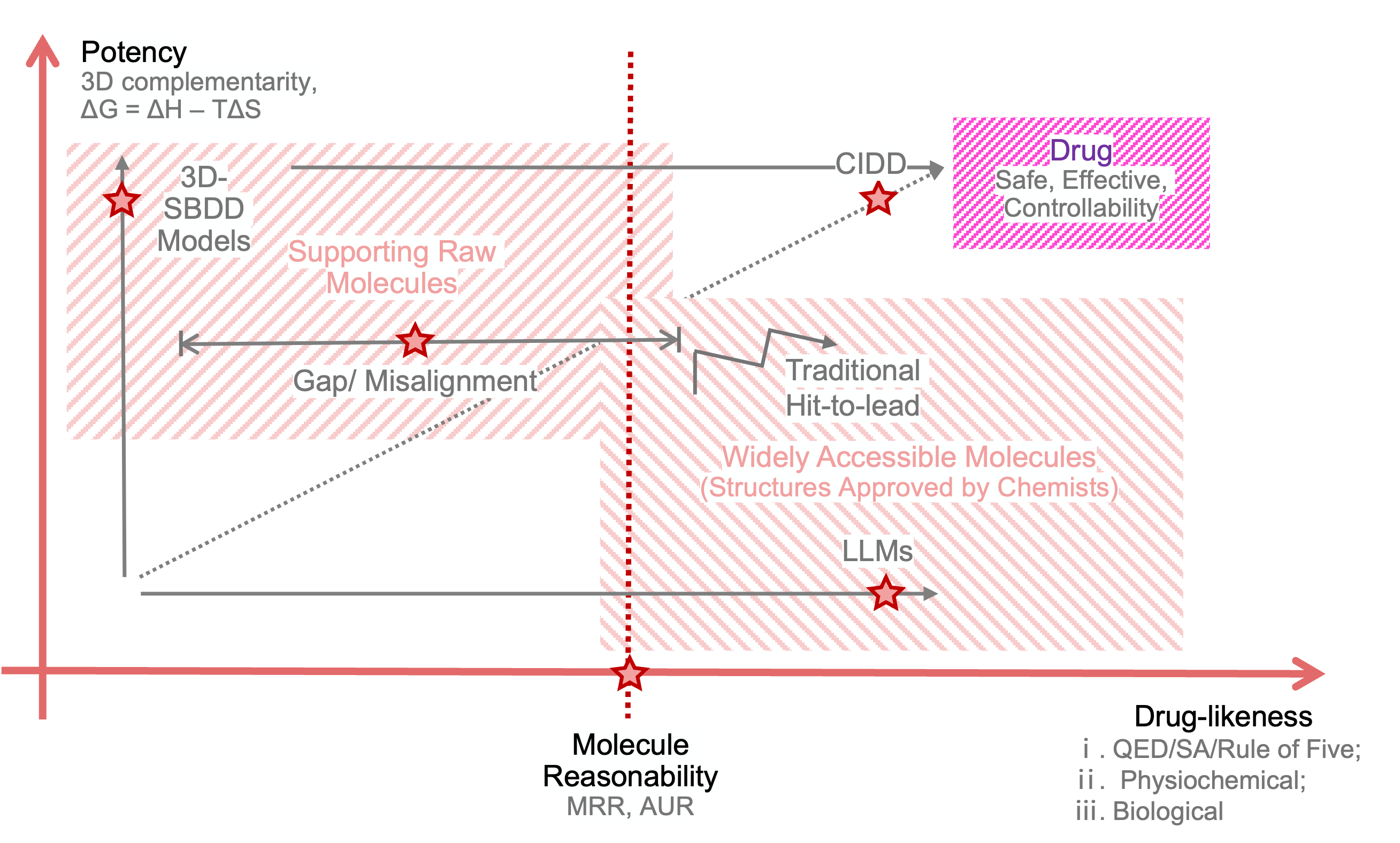}}
\caption{\textbf{The CIDD framework generates molecules with both exceptional potency and favorable drug-like properties.} While 3D-SBDD models can produce molecules with strong 3D complementarity to target structures, these outputs often lack drug-like characteristics and are chemically unreasonable. In contrast, LLMs excel at generating chemically reasonable small molecules with good drug-likeness. CIDD effectively integrates the strengths of both approaches. Furthermore, the significant gap—or misalignment—between raw supporting molecules and widely accessible molecules was identifies and quantifies using novel metrics introduced in this work.}
\label{fig:workflow}
\end{center}
\end{figure*}

\subsection{Problem Identification}
In exploring how AI models could be applied to various stages of the drug discovery process, it is often the case that AI applications are directly copied from the key steps of the traditional drug development workflow without a thorough analysis of the content and meaning of the tasks. However, effectively applying AI in pharmaceutical science requires a focus on the core issues AI needs to solve, as well as an analysis of the new challenges and problems that arise as AI addresses these tasks.

Through research into the outputs of various 3D-SBDD models, we find that these outputs differ significantly from the traditional starting point of the ``hit-to-lead” process. Specifically, the drug-likeness of these outputs is problematic, and the more challenging issue lies in the fact that many of these molecules are not even reasonable. Therefore, the problem faced by current 3D-SBDD models is not simply optimizing drug-likeness in the traditional ``hit-to-lead" sense, but rather addressing the broader challenge of transforming unreasonable molecules into reasonable and drug-like candidates.

We believe that only by fully describing and demonstrating the gap between 3D-SBDD model results and compound structures widely accepted by (medicinal) chemists can we effectively bridge it, ultimately delivering practical benefits to AIDD and guiding its development in the right direction. In this work, by defining ``MRR” and ``AUR”, we effectively automate the identification of errors in model outputs regarding reasonability, specifically distinguishing unreasonable 3D-SBDD model outputs from molecules approved by (medicinal) chemists. Moreover, these two novel metrics are simple, relying solely on molecular topology, and do not depend on 3D conformation prediction or molecular property prediction. This is an important contribution, as it concretely highlights the misalignment between AI model outputs and molecules that can be broadly accepted and applied in pharmaceutical research. Furthermore, our algorithms can quantitatively measure this reasonability gap.

\subsection{Problem Solving}
Drug discovery aims to identify therapeutic molecules that satisfy all criteria for safety, efficacy, and controllable quality, including at least two dimensions: potency and drug-likeness. Potency is primarily influenced by the degree of 3D complementarity between designed small molecules and their intended target structures, while drug-likeness is determined by the general physicochemical or biological properties of the small molecules. Potency is often roughly evaluated using docking scores, such as the Vina score, while metrics such as QED and SA are used to estimate drug-likeness. In this paper, we argue that a molecule must at least be ``reasonable” to be considered drug-like, and drug-likeness can, in turn, be broken down into simple empirical rules, individual physicochemical metrics, and biological metrics (see Appendix \ref{sec: algorithm}, \ref{sec:qkiprop} and \ref{sec: three level qikprop}). We have successfully ensured that all detailed aspects of drug-likeness requirements are simultaneously and intelligently addressed within our framework.

In practical medicinal chemistry, the ``hit-to-lead" process follows a problem-solving strategy, where each issue is addressed individually. It is a resource- and time-intensive process, often resulting in a trade-off between improving drug-likeness and decreasing potency, and vice versa. Therefore, this paper achieves a significant and challenging task: achieving substantial improvements in drug-likeness while maintaining potency.

Moreover, as previously discussed, challenges related to drug-likeness manifest in diverse ways depending on the specific characteristics of the molecule. For instance, certain compounds may exhibit inadequate water solubility, whereas others might face limitations in oral bio-availability. These variations underscore the necessity of tailoring strategies to enhance drug-likeness to the unique properties of each small molecule. Achieving this level of customization demands a thorough evaluation of multiple factors to develop the most effective optimization strategy—a process that traditionally requires extensive expertise, experience, and iterative experimentation by medicinal chemists. In this study, however, we have revolutionized this approach by harnessing the integrated knowledge and reasoning capabilities of LLMs to automate the process, marking a significant advancement in the field.

\section{Conclusion and Future Works}

% In this paper, we introduced the \textbf{CIDD} framework, which overcomes key limitations of traditional SBDD models by integrating advanced 3D-SBDD models—capable of capturing molecular interactions—with Large Language Models, which excel in understanding drug requirements. Our experimental evaluations on the CrossDocked2020 dataset demonstrate that \textbf{CIDD} significantly outperforms SOTA methods, achieving a higher success ratio, defined by interaction ability and drug-likeness. This work pioneers the collaboration between domain-specific models and LLMs, showcasing its effectiveness in enhancing molecular generation. Beyond this, the proposed framework can be extended to other critical stages of drug discovery, including \textbf{target identification}, \textbf{hit-to-lead optimization}, \textbf{preclinical toxicity evaluation}, and \textbf{synthetic pathway planning}, which we identify as future works. We believe that the collaboration between task-specific machine learning models and LLMs has the potential to revolutionize drug discovery, enabling a more efficient, interpretable, and impactful pipeline.

In this paper, we introduced the \textbf{CIDD} framework, which addresses key limitations of traditional SBDD by integrating advanced 3D-SBDD models—capable of capturing molecular interactions—with LLMs, which excel in understanding drug-likeness requirements. Our evaluations on the CrossDocked2020 dataset show that \textbf{CIDD} outperforms SOTA methods, achieving a higher success ratio defined by interaction ability and drug-likeness. This work pioneers the collaboration between domain-specific models and LLMs, demonstrating its effectiveness in molecular generation. Beyond this, \textbf{CIDD} can extend to other critical drug discovery stages, including \textbf{target identification}, \textbf{hit-to-lead optimization}, \textbf{preclinical toxicity evaluation}, and \textbf{synthetic pathway planning}, which we identify as future works. We believe the collaboration between task-specific machine learning models and LLMs can revolutionize drug discovery, enabling a more efficient, interpretable, and impactful pipeline.

\bibliography{paper}
\bibliographystyle{icml2025}

%%%%%%%%%%%%%%%%%%%%%%%%%%%%%%%%%%%%%%%%%%%%%%%%%%%%%%%%%%%%%%%%%%%%%%%%%%%%%%%
%%%%%%%%%%%%%%%%%%%%%%%%%%%%%%%%%%%%%%%%%%%%%%%%%%%%%%%%%%%%%%%%%%%%%%%%%%%%%%%
% APPENDIX
%%%%%%%%%%%%%%%%%%%%%%%%%%%%%%%%%%%%%%%%%%%%%%%%%%%%%%%%%%%%%%%%%%%%%%%%%%%%%%%
%%%%%%%%%%%%%%%%%%%%%%%%%%%%%%%%%%%%%%%%%%%%%%%%%%%%%%%%%%%%%%%%%%%%%%%%%%%%%%%
\newpage
\appendix
\onecolumn

\section{Detailed Prompts and Responses for LEDD} \label{sec:prompts}

In this section, we present the detailed workflow of the CIDD framework, including the prompts and example responses for each module.

Figure \ref{fig:workflow} illustrates the complete drug design pipeline. The Interaction Module first identifies key fragments within the supporting molecule that interact with the protein pocket. This information is then utilized by the Design Module, which devises strategies to replace uncommon or unfavorable fragments while preserving crucial interactions. Once a new molecule is designed, the Evaluation Phase within the Design Module assesses its viability. Finally, the Reflection Module analyzes the design process and outcomes, highlighting both strengths and areas for improvement.

\begin{figure}[H]
\begin{center}
\centerline{\includegraphics[width=\textwidth]{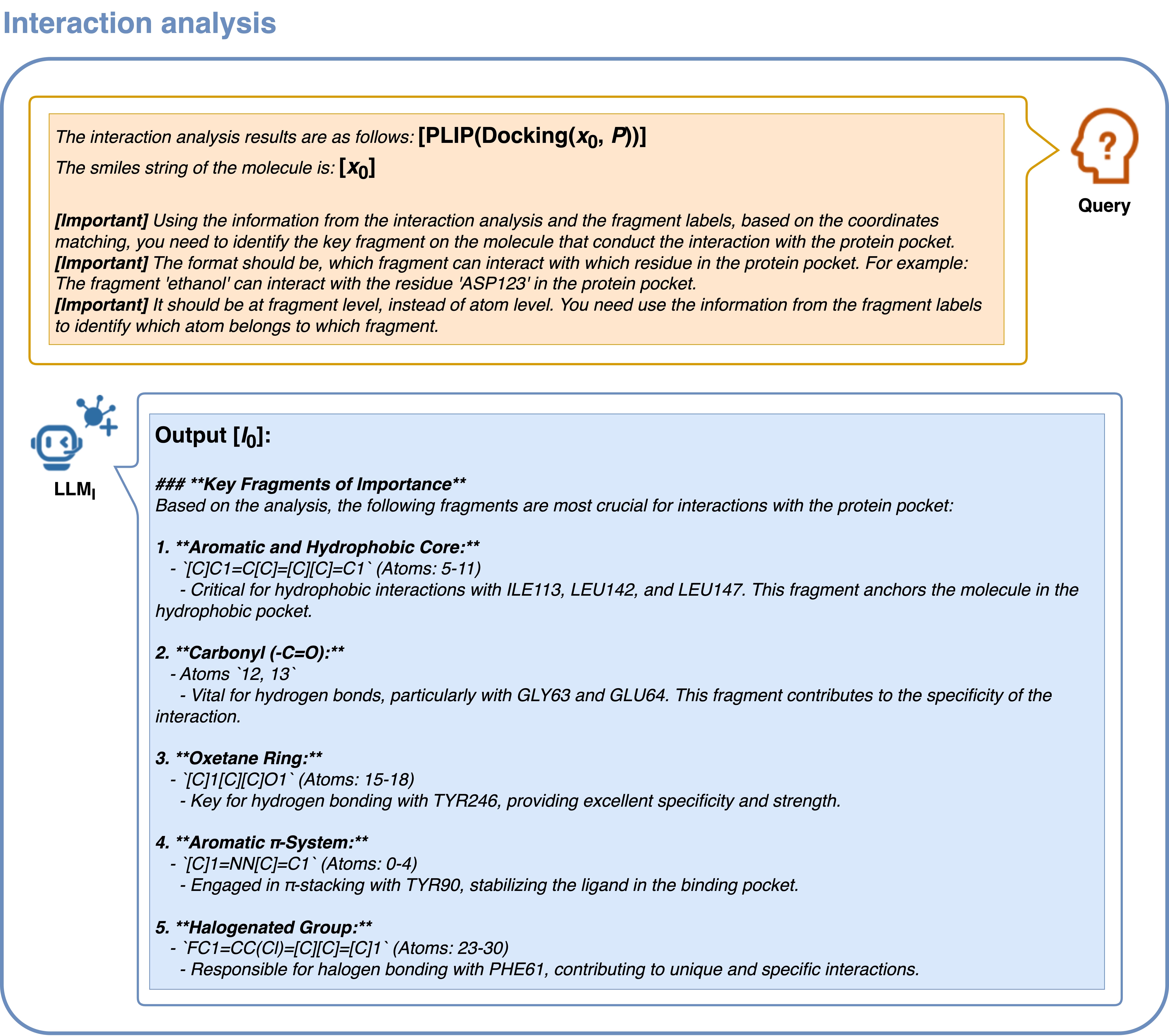}}
\caption{Interaction analysis module}
\label{fig: interaction response}
\end{center}
\end{figure}

Figure~\ref{fig: interaction response} presents the prompt and example response for the Interaction Analysis Module.

\begin{figure}[H]
\begin{center}
\centerline{\includegraphics[width=\textwidth]{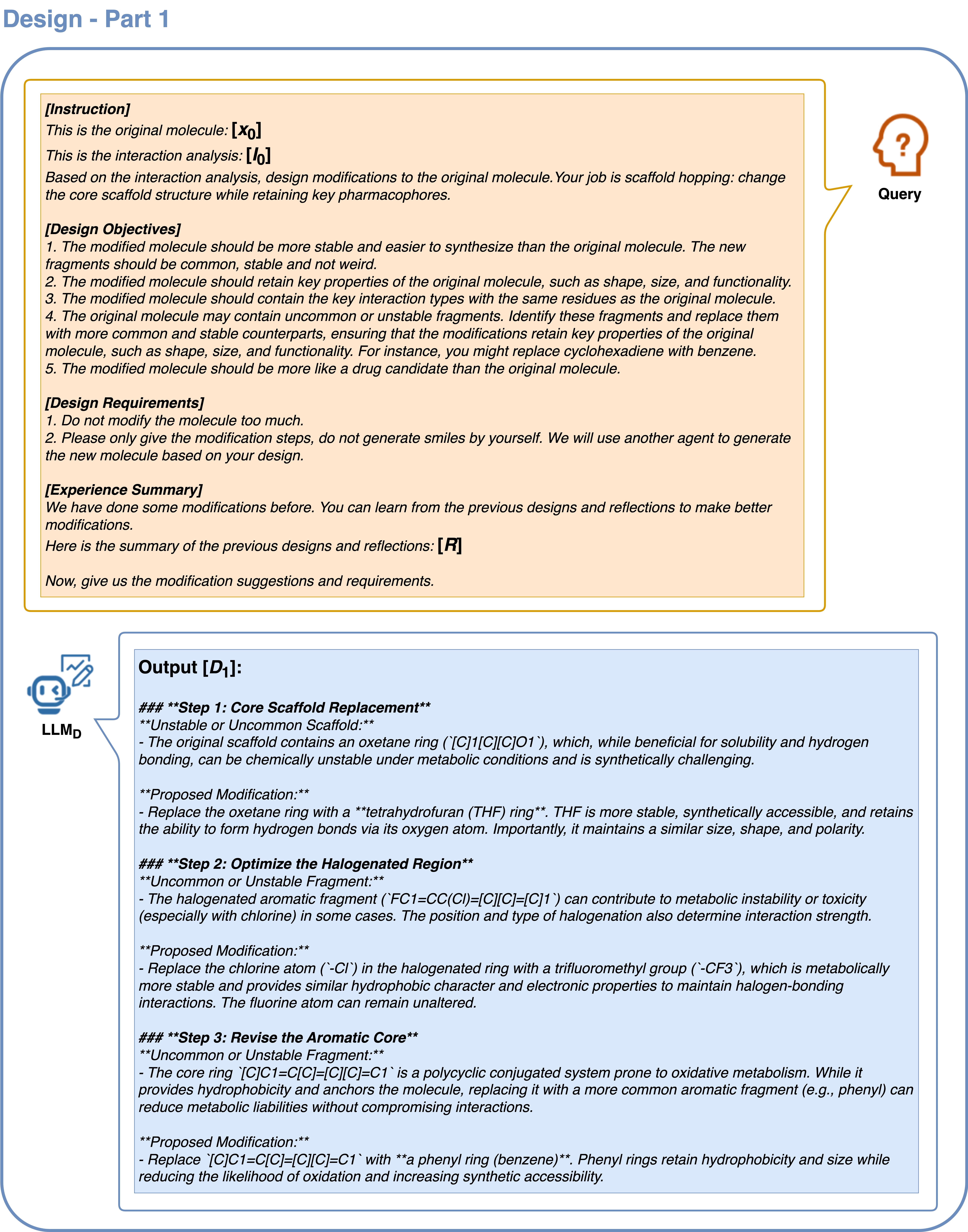}}
\caption{Design Module}
\label{fig: design response 1}
\end{center}
\end{figure}

\begin{figure}[H]
\begin{center}
\centerline{\includegraphics[width=\textwidth]{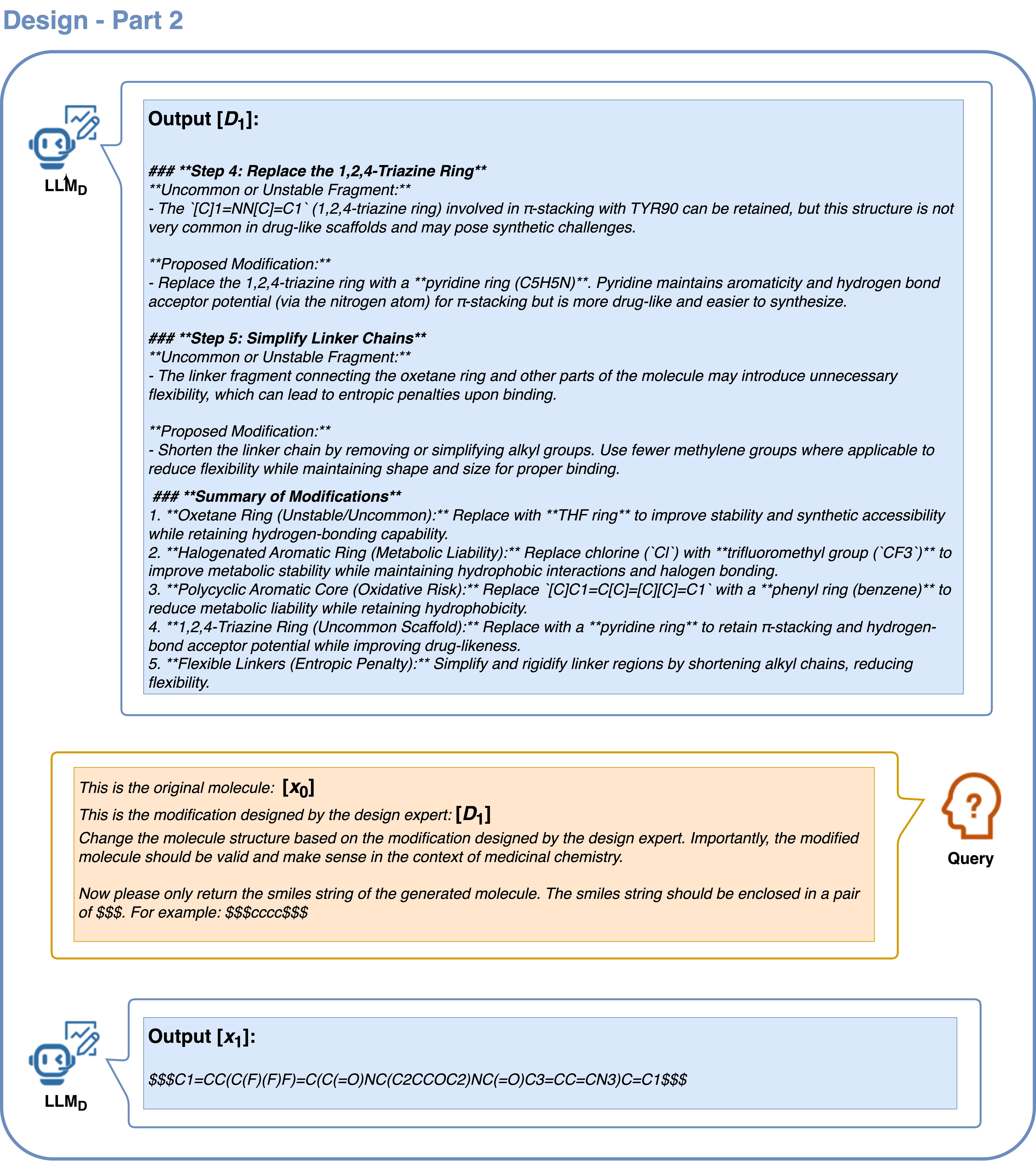}}
\caption{Design Module}
\label{fig: design response 2}
\end{center}
\end{figure}

Figures~\ref{fig: design response 1} and~\ref{fig: design response 2} display the prompt and example response for the Design Module.

\begin{figure}[H]
\begin{center}
\centerline{\includegraphics[width=\textwidth]{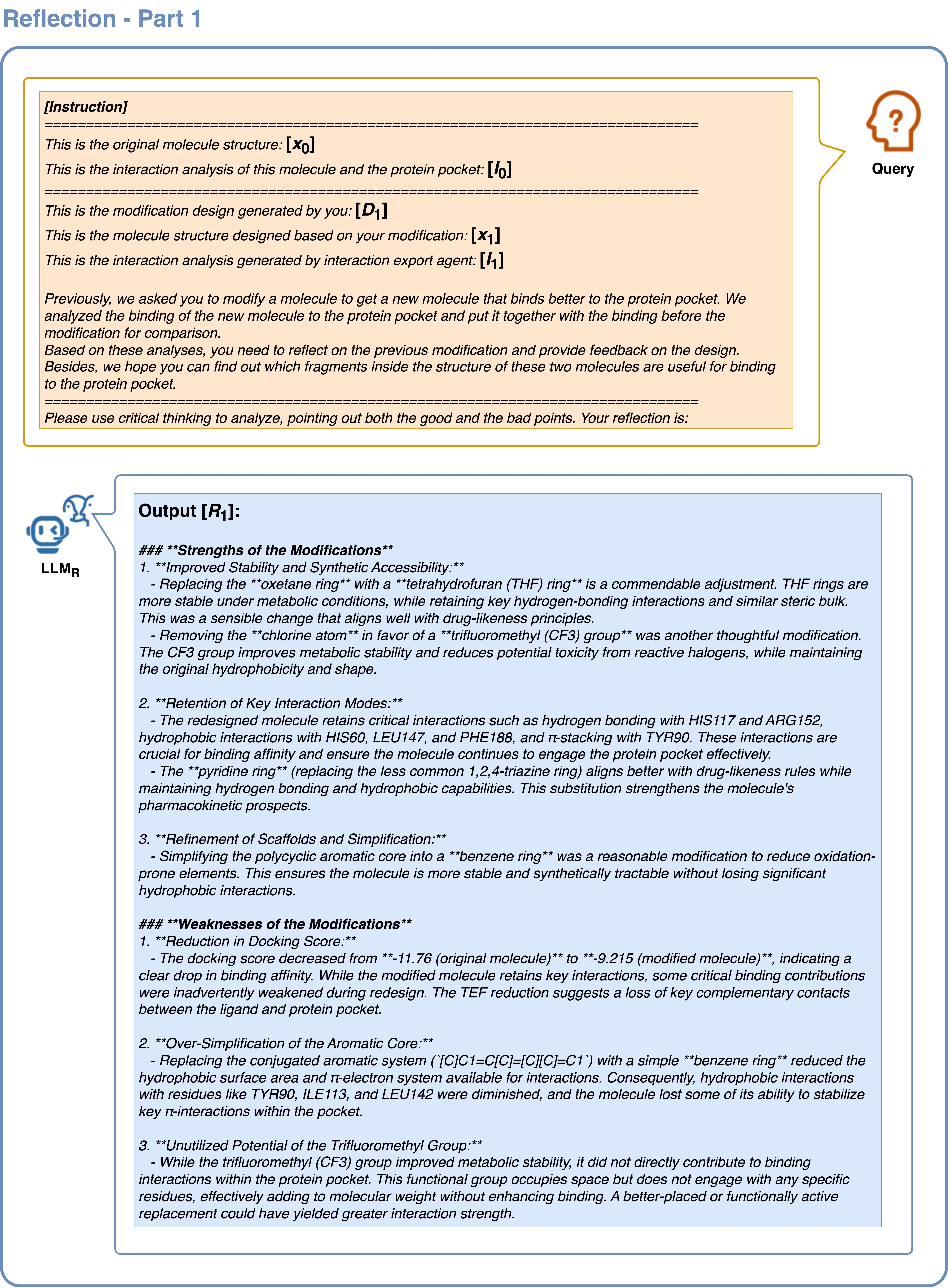}}
\caption{Reflection Module}
\label{fig: reflection response 1}
\end{center}
\end{figure}

\begin{figure}[H]
\begin{center}
\centerline{\includegraphics[width=\textwidth]{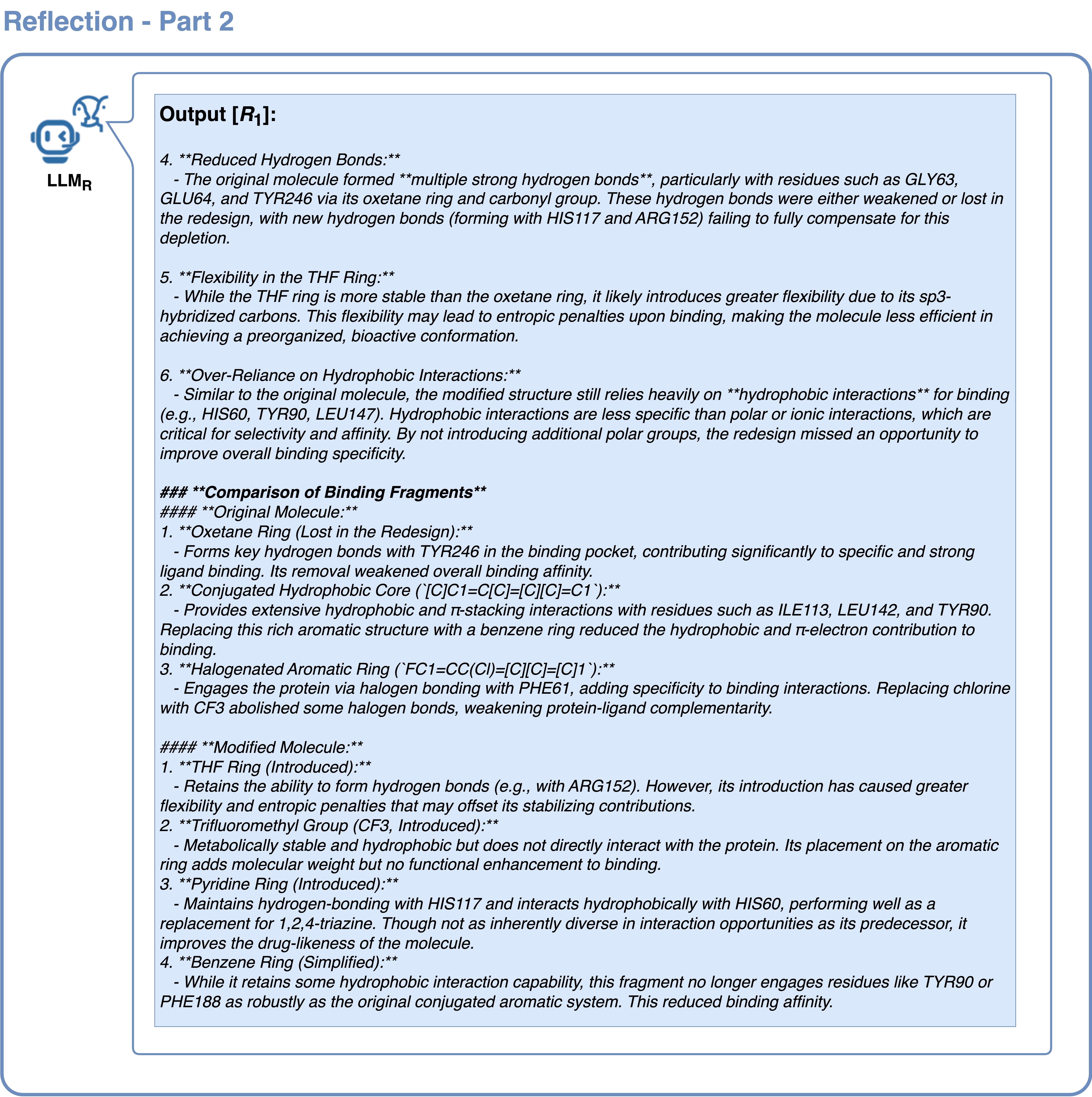}}
\caption{Reflection Module}
\label{fig: reflection response 2}
\end{center}
\end{figure}

\begin{figure}[H]
\begin{center}
\centerline{\includegraphics[width=\textwidth]{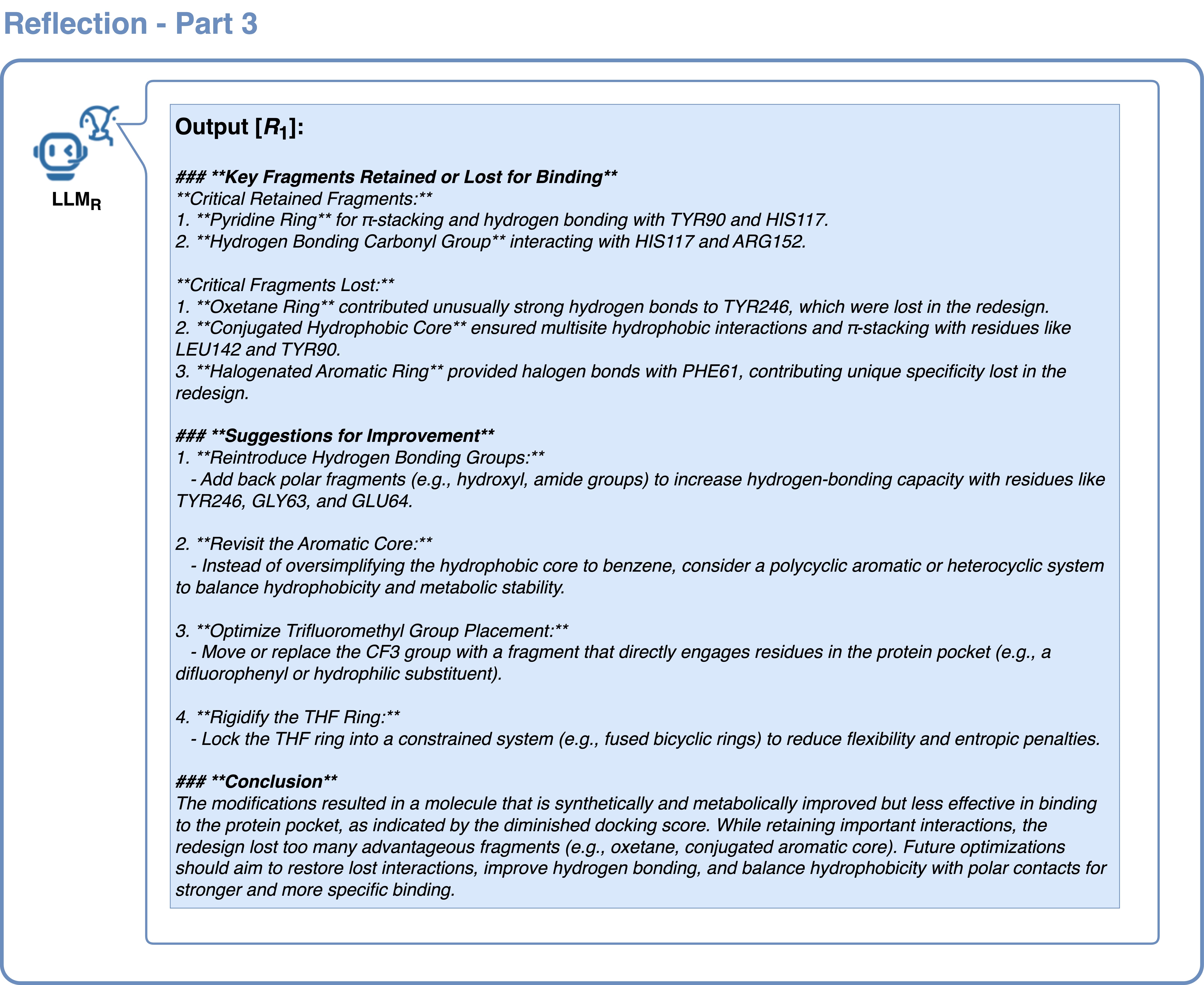}}
\caption{Reflection Module}
\label{fig: reflection response 3}
\end{center}
\end{figure}

Figures~\ref{fig: reflection response 1}, \ref{fig: reflection response 2}, and \ref{fig: reflection response 3} illustrate the prompt and example responses for the Reflection Module.

\begin{figure}[H]
\begin{center}
\centerline{\includegraphics[width=0.9\textwidth]{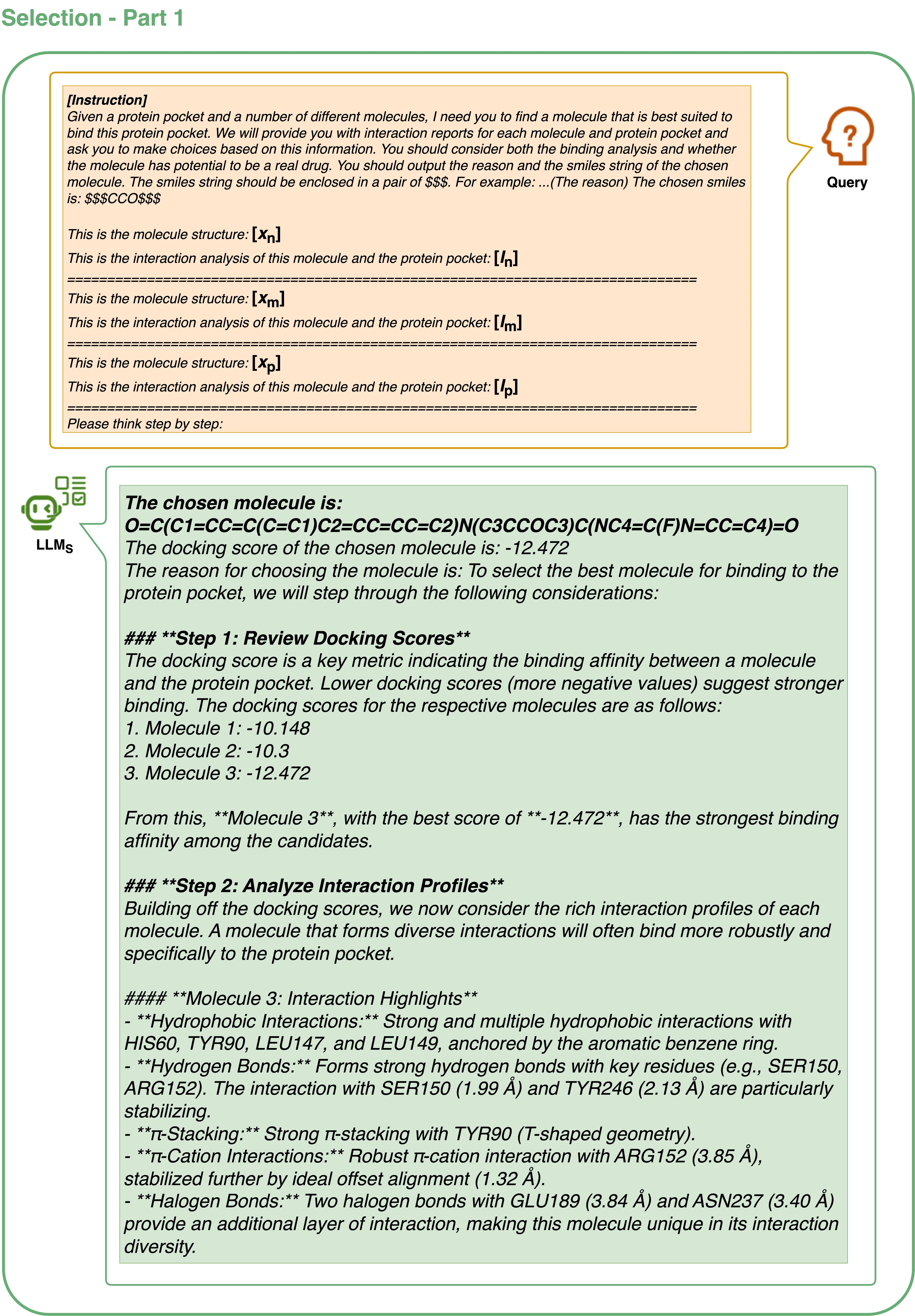}}
\caption{Selection Module}
\label{fig: selection response 1}
\end{center}
\end{figure}

\begin{figure}[H]
\begin{center}
\centerline{\includegraphics[width=0.9\textwidth]{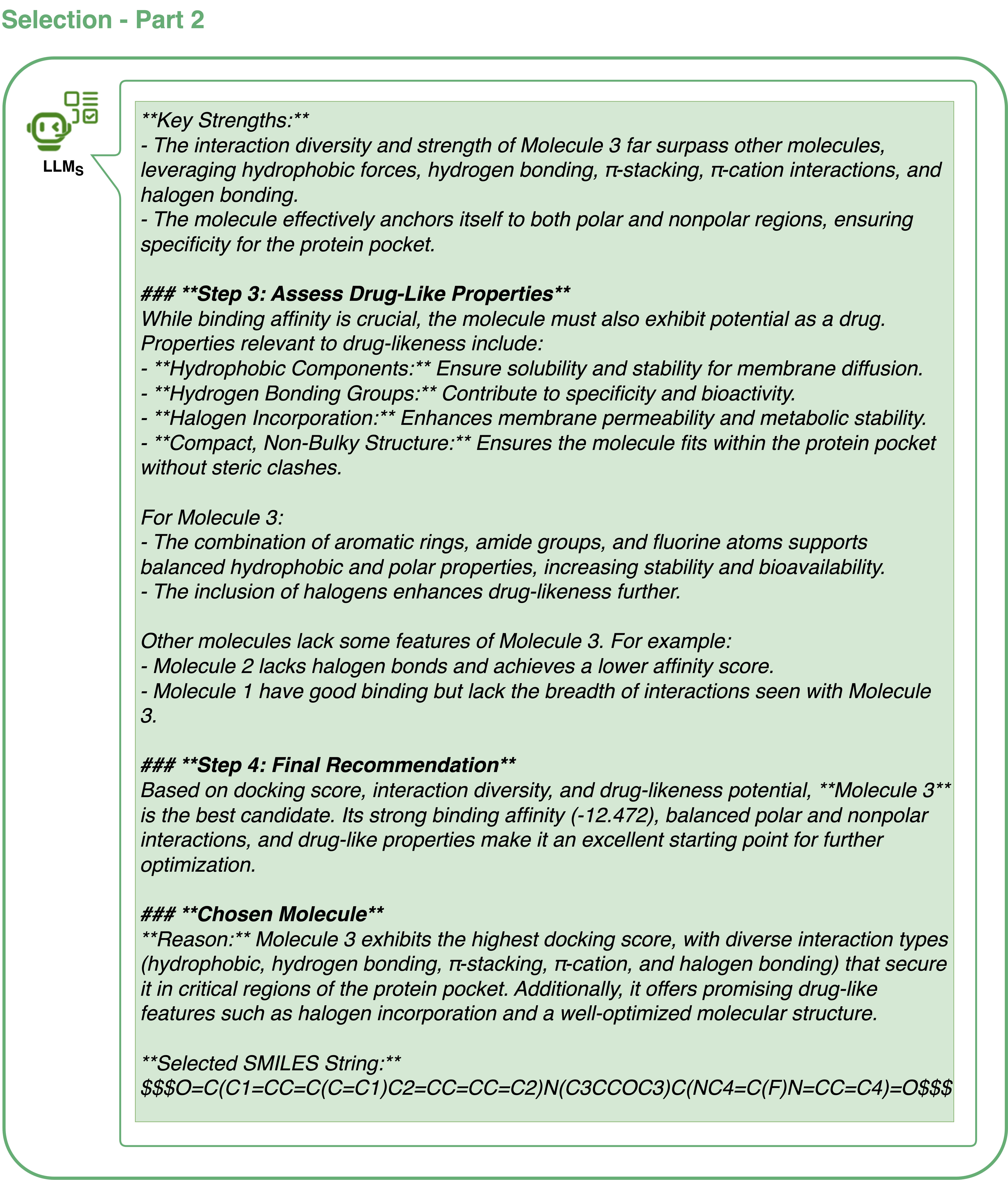}}
\caption{Selection Module}
\label{fig: selection response 2}
\end{center}
\end{figure}

Figures~\ref{fig: selection response 1} and~\ref{fig: selection response 2} show the prompt and example response for the Selection Module.

\begin{figure}[ht]
\begin{center}
\includegraphics[width=\linewidth]{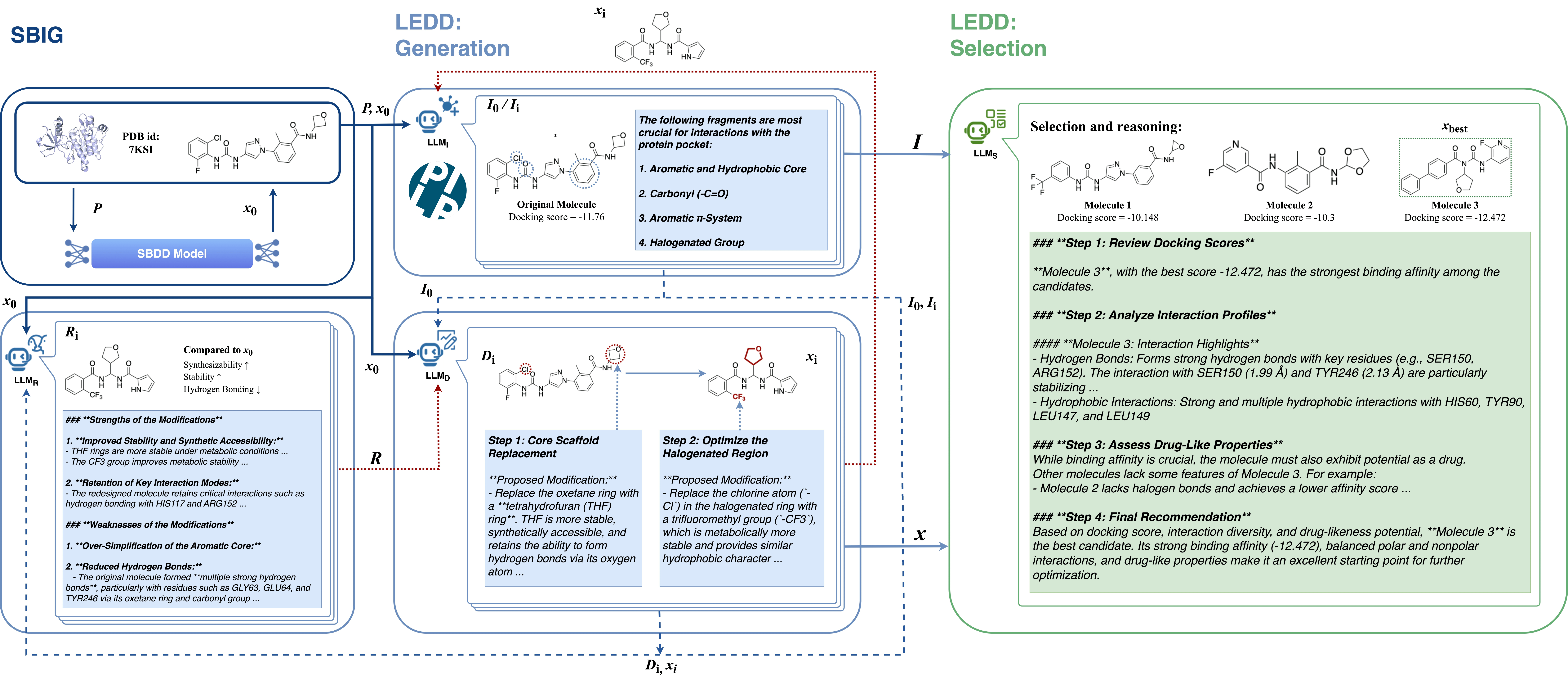}
%\includegraphics[width=15cm]{fig/workflow_compressed.jpeg}
%\resizebox{\textwidth}{!}{\includegraphics{fig/workflow_compressed.jpeg}}
\caption{Workflow of CIDD framework}
\label{fig:workflow}
\end{center}
\end{figure}

\section{Algorithm for MRR and AUR}
\label{sec: algorithm}

The complete calculation process for assessing the reasonability of a molecule is outlined in Algorithm~\ref{alg:reasonability}.

\begin{algorithm}[H]
\caption{Evaluation of Molecular Reasonability}
\label{alg:reasonability}
\KwIn{Molecule object (\textit{mol})}
\KwOut{Molecular Reasonability (\textit{MRR}) and Atom Unreasonable Ratio (\textit{AUR})}

\textbf{Step 1: Detect Carbonyl and Imine Group Carbons} \\
Initialize an empty list for \textit{carbonyl/imine carbons}. \\
\ForEach{bond in \textit{mol}}{
    \If{bond is double and one atom is carbon, the other is oxygen or nitrogen}{
        Record the carbon atom in \textit{carbonyl/imine groups}.
    }
}

\textbf{Step 2: Identification of Ring Systems} \\
Identify all ring structures and their corresponding atom indices within \textit{mol}. \\
Calculate the number of atoms in each ring. \\
\ForEach{ring in the molecule}{
    \If{the ring shares one or more atoms with another ring}{
        Group the connected rings into a single \textit{ring system}.
    }
}

\textbf{Step 3: Evaluation of Molecular Reasonability} \\
    Exclude any atoms previously identified as part of carbonyl or imine groups. \\
    Classify the remaining carbon atoms in each ring system as follows:
\begin{itemize}
    \setlength{\itemsep}{-6pt}  % Reduce space between items
    \item \textit{sp\textsuperscript{2} hybridized}: Aromatic or unsaturated carbons.
    \item \textit{Non-sp\textsuperscript{2} hybridized}: Saturated carbons.
\end{itemize}

\ForEach{ring system in the \textit{ring systems}}{
    \If{the ring system contains multiple rings and all carbon atoms are \textit{non-sp\textsuperscript{2}}}{
        Mark the molecule as unreasonable. \\
        Add the atoms to the unreasonable atom list.
    }
}

\ForEach{ring system in the \textit{remaining ring systems}}{
    \ForEach{ring in the ring system}{
        \If{all carbon atoms within the ring are consistent in hybridization (either all \textit{sp\textsuperscript{2}} or all \textit{non-sp\textsuperscript{2}})}{
            Mark the ring as reasonable.
        }
        \Else{
            Add the ring to the \textit{remaining ring list}.
        }
    }
}

\While{the \textit{remaining ring list} is not empty}{
    \ForEach{ring in the \textit{remaining ring list}}{
        Exclude atoms that have already been classified as reasonable. \\
        \If{all remaining carbon atoms are consistent in hybridization (either all \textit{sp\textsuperscript{2}} or all \textit{non-sp\textsuperscript{2}})}{
            Mark the ring as reasonable.
        }
    }

    \If{no new reasonable rings are identified}{
        Mark the molecule as unreasonable. \\
        Add the carbon atoms in the remaining rings to the unreasonable atom list. \\
        \textbf{Exit the loop.}
    }
}

Calculate \textit{AUR} as the ratio of unreasonable atom count to the total ring atom count.

\textbf{Return} \textit{MRR} and \textit{AUR}.
\end{algorithm}

\section{QikProp properties}\label{sec:qkiprop}

The full set of properties used for the QikProp pass ratio analysis is presented in Table \ref{tab:qikprop_properties}.

The QikProp filter applied in the main text incorporates a comprehensive range of criteria provided by QikProp, including "\#stars", "\#amine", "\#amidine", "\#acid", "\#amide", "\#rotor", "\#rtvFG", "mol\_MW", "dipole", "SASA", "FOSA", "FISA", "PISA", "WPSA", "volume", "donorHB", "accptHB", "dip$^2/V$", "ACxDN$^.5/$SA", "glob", "QPpolrz", "QPlogPC16", "QPlogPoct", "QPlogPw", "QPlogPo/w", "QPlogS", "CIQPlogS", "QPPCaco", "QPlogBB", "QPPMDCK", "QPlogKp", "IP(eV)", "EA(eV)", "\#metab", "QPlogKhsa", "PercentHumanOralAbsorption", "SAFluorine", "SAamideO", "PSA", "\#NandO", and "RuleOfThree".

\begin{table}[h!]
\centering
\caption{QikProp Properties and Descriptors}
\label{tab:qikprop_properties}
\resizebox{\textwidth}{!}{
\begin{tabular}{|l|p{12cm}|l|}
\hline
\textbf{Property or Descriptor} & \textbf{Description} & \textbf{Range or Recommended Values} \\ \hline
Molecule name & The molecule's identifier derived from the title line in the input structure file. If no title is provided, the file name is used. & \\ \hline
\#stars & Count of descriptors or properties falling outside the 95\% range for known drugs. A higher count indicates reduced drug-likeness. & 0 -- 5 \\ \hline
\#amine & Total non-conjugated amine groups present in the molecule. & 0 -- 1 \\ \hline
\#amidine & Number of amidine or guanidine functional groups in the structure. & 0 \\ \hline
\#acid & Quantity of carboxylic acid groups in the molecule. & 0 -- 1 \\ \hline
\#amide & Count of non-conjugated amide groups. & 0 -- 1 \\ \hline
\#rotor & Number of rotatable bonds that are neither trivial nor sterically hindered. & 0 -- 15 \\ \hline
\#rtvFG & Total reactive functional groups present in the molecule, potentially affecting stability or toxicity. & 0 -- 2 \\ \hline
% CNS & Predicted activity on the central nervous system, ranging from -2 (inactive) to +2 (active). & -2 to +2 \\ \hline
mol\_MW & Molecular weight of the compound. & 130.0 -- 725.0 \\ \hline
Dipole & Calculated dipole moment of the molecule in Debye units. & 1.0 -- 12.5 \\ \hline
SASA & Solvent-accessible surface area (SASA) in square angstroms, measured with a probe of 1.4 Å radius. & 300.0 -- 1000.0 \\ \hline
FOSA & Hydrophobic part of the SASA, representing saturated carbon and attached hydrogen atoms. & 0.0 -- 750.0 \\ \hline
FISA & Hydrophilic fraction of the SASA, encompassing polar atoms like nitrogen and oxygen. & 7.0 -- 330.0 \\ \hline
PISA & SASA component attributable to $\pi$-systems. & 0.0 -- 450.0 \\ \hline
WPSA & Weakly polar component of the SASA, including atoms like halogens, phosphorus, and sulfur. & 0.0 -- 175.0 \\ \hline
Volume & Total solvent-accessible volume in cubic angstroms, determined with a 1.4 Å radius probe. & 500.0 -- 2000.0 \\ \hline
donorHB & Estimated number of hydrogen bonds donated to water in solution. & 0.0 -- 6.0 \\ \hline
accptHB & Estimated number of hydrogen bonds accepted from water. & 2.0 -- 20.0 \\ \hline
Dip$^2$/V & Dipole moment squared divided by molecular volume, a key factor in solvation energy. & 0.0 -- 0.13 \\ \hline
ACxDN$^{0.5}$/SA & Cohesive interaction index in solids based on molecular properties. & 0.0 -- 0.05 \\ \hline
glob & Descriptor measuring how close the shape of a molecule is to a sphere. & 0.75 -- 0.95 \\ \hline
QPpolrz & Predicted molecular polarizability in cubic angstroms. & 13.0 -- 70.0 \\ \hline
QPlogPC16 & Predicted partition coefficient between hexadecane and gas phases. & 4.0 -- 18.0 \\ \hline
QPlogPoct & Predicted partition coefficient between octanol and gas phases. & 8.0 -- 35.0 \\ \hline
QPlogPw & Predicted partition coefficient between water and gas phases. & 4.0 -- 45.0 \\ \hline
QPlogPo/w & Predicted partition coefficient between octanol and water phases. & -2.0 -- 6.5 \\ \hline
QPlogS & Predicted solubility of the molecule in water (log S, in mol/L). & -6.5 -- 0.5 \\ \hline
CIQPlogS & Conformation-independent prediction of water solubility (log S). & -6.5 -- 0.5 \\ \hline
% QPlogHERG & Predicted IC50 for HERG potassium channel inhibition. & Concern below -5 \\ \hline
QPPCaco & Predicted permeability through Caco-2 cells, in nm/s. & $<$25 poor, $>$500 great \\ \hline
QPlogBB & Predicted partition coefficient for brain/blood. & -3.0 -- 1.2 \\ \hline
QPPMDCK & Predicted permeability through MDCK cells, in nm/s. & $<$25 poor, $>$500 great \\ \hline
QPlogKp & Predicted skin permeability (log Kp). & -8.0 -- -1.0 \\ \hline
IP(eV) & Ionization potential calculated using PM3. & 7.9 -- 10.5 \\ \hline
EA(eV) & Electron affinity calculated using PM3. & -0.9 -- 1.7 \\ \hline
\#metab & Predicted number of possible metabolic reactions. & 1 -- 8 \\ \hline
QPlogKhsa & Predicted binding affinity to human serum albumin. & -1.5 -- 1.5 \\ \hline
HumanOralAbsorption & Qualitative assessment of oral absorption: 1 (low), 2 (medium), or 3 (high). & \\ \hline
PercentHumanOralAbsorption & Quantitative prediction of oral absorption percentage. & $>$80\% high, $<$25\% poor \\ \hline
SAFluorine & Solvent-accessible fluorine surface area. & 0.0 -- 100.0 \\ \hline
SAamideO & Solvent-accessible surface area of amide oxygen atoms. & 0.0 -- 35.0 \\ \hline
PSA & Polar surface area, calculated for nitrogen, oxygen, and carbonyl groups. & 7.0 -- 200.0 \\ \hline
\#NandO & Total count of nitrogen and oxygen atoms. & 2 -- 15 \\ \hline
RuleOfFive & Number of Lipinski’s Rule of Five violations. & Max 4 \\ \hline
RuleOfThree & Number of Jorgensen’s Rule of Three violations. & Max 3 \\ \hline
\#ringatoms & Count of atoms within molecular rings. & \\ \hline
\#in34 & Number of atoms in 3- or 4-membered rings. & \\ \hline
\#in56 & Number of atoms in 5- or 6-membered rings. & \\ \hline
\#noncon & Number of ring atoms unable to form conjugated aromatic systems. & \\ \hline
\#nonHatm & Count of heavy (non-hydrogen) atoms in the structure. & \\ \hline
Jm & Predicted maximum transdermal transport rate ($\mu$g cm$^{-2}$ hr$^{-1}$). & \\ \hline
\end{tabular}%
}
\end{table}

\section{More Experiment Results}\label{sec: three level qikprop}

Based on the different criteria presented in Table \ref{tab:qikprop_properties}, we provide additional pass ratio results in Table \ref{tab:all qikprop}. 

Filter 1 is identical to the QikProp filter used in the main text.

Filter 2 removes some non-essential properties and focuses on well-defined physicochemical properties, including "\#rtvFG", "QPlogS", "QPlogPo/w", "mol\_MW", "dipole", "SASA", "FOSA", "FISA", "IP(eV)", "EA(eV)", "\#metab", "PercentHumanOralAbsorption", and "PSA".

Filter 3 assesses molecular compliance with the "RuleOfFive" criterion. However, instead of allowing up to four violations as typically recommended, this filter adopts a stricter definition, considering only molecules that fully comply (i.e., setting the maximum allowable violations to zero).

\begin{table}[h]
    \centering
    \caption{QikProp results for different methods with and without CIDD}
    \vspace{2pt}
    \begin{tabular}{lccc}
        \toprule
        Method      & Filter 1 & Filter 2 & Filter 3 \\
        \midrule
        \multicolumn{4}{l}{\textbf{Pocket2Mol}} \\
        Original    & 29.58\%  & 51.52\%  & 89.58\%  \\
        CIDD        & 56.97\%  & 75.64\%  & 92.24\%  \\
        \midrule
        \multicolumn{4}{l}{\textbf{TargetDiff}} \\
        Original    & 26.32\%  & 48.20\%  & 69.47\%  \\
        CIDD        & 53.37\%  & 75.60\%  & 81.85\%  \\
        \midrule
        \multicolumn{4}{l}{\textbf{DecompDiff}} \\
        Original    & 29.04\%  & 53.96\%  & 55.14\%  \\
        CIDD        & 37.54\%  & 68.48\%  & 65.64\%  \\
        \midrule
        \multicolumn{4}{l}{\textbf{MolCRAFT}} \\
        Original    & 22.37\%  & 43.52\%  & 66.45\%  \\
        CIDD        & 35.22\%  & 63.23\%  & 74.09\%  \\
        \bottomrule
    \end{tabular}
    
    \label{tab:all qikprop}
\end{table}

\section{More cases}
\label{sec: more cases}

More generated molecules from CIDD are presented below. For each case, we display the initial supporting molecule derived from 3D-SBDD models alongside the final designed molecules produced by CIDD.

\begin{figure}[H]
\begin{center}
\centerline{\includegraphics[width=\textwidth]{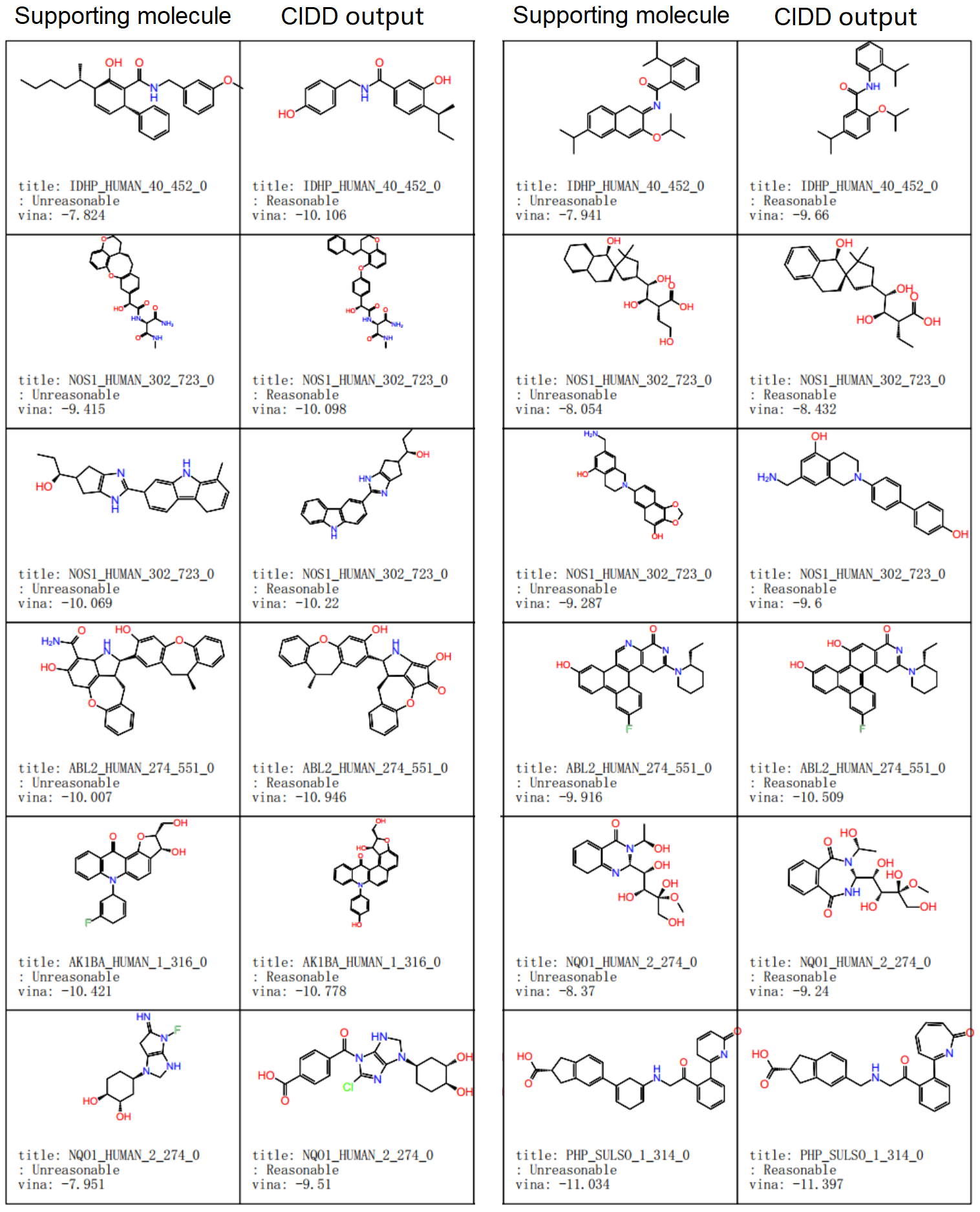}}
% \caption{Interaction analysis module}
\label{fig: case1}
\end{center}
\end{figure}

\begin{figure}[H]
\begin{center}
\centerline{\includegraphics[width=\textwidth]{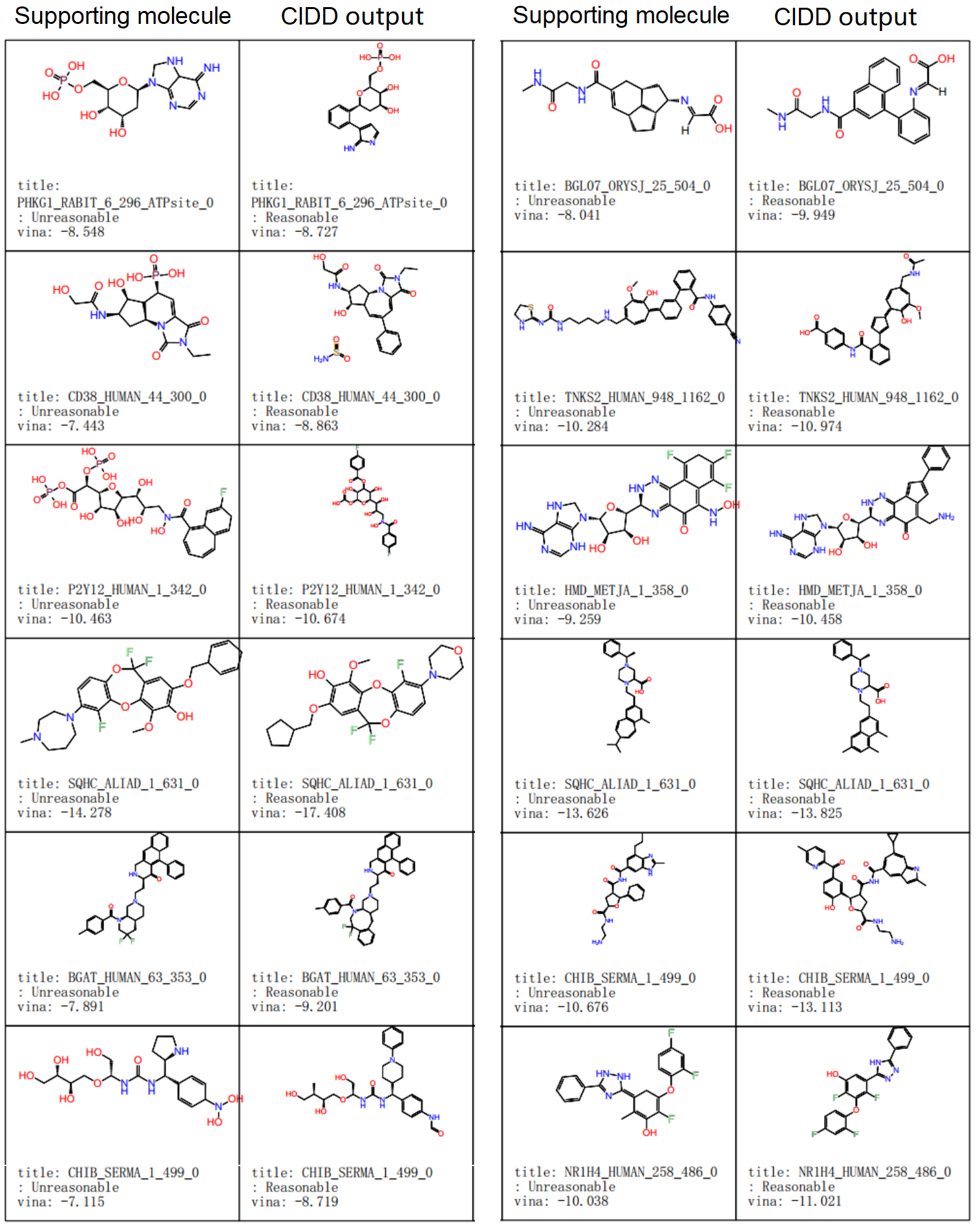}}
% \caption{Interaction analysis module}
\label{fig: case2}
\end{center}
\end{figure}

\begin{figure}[H]
\begin{center}
\centerline{\includegraphics[width=\textwidth]{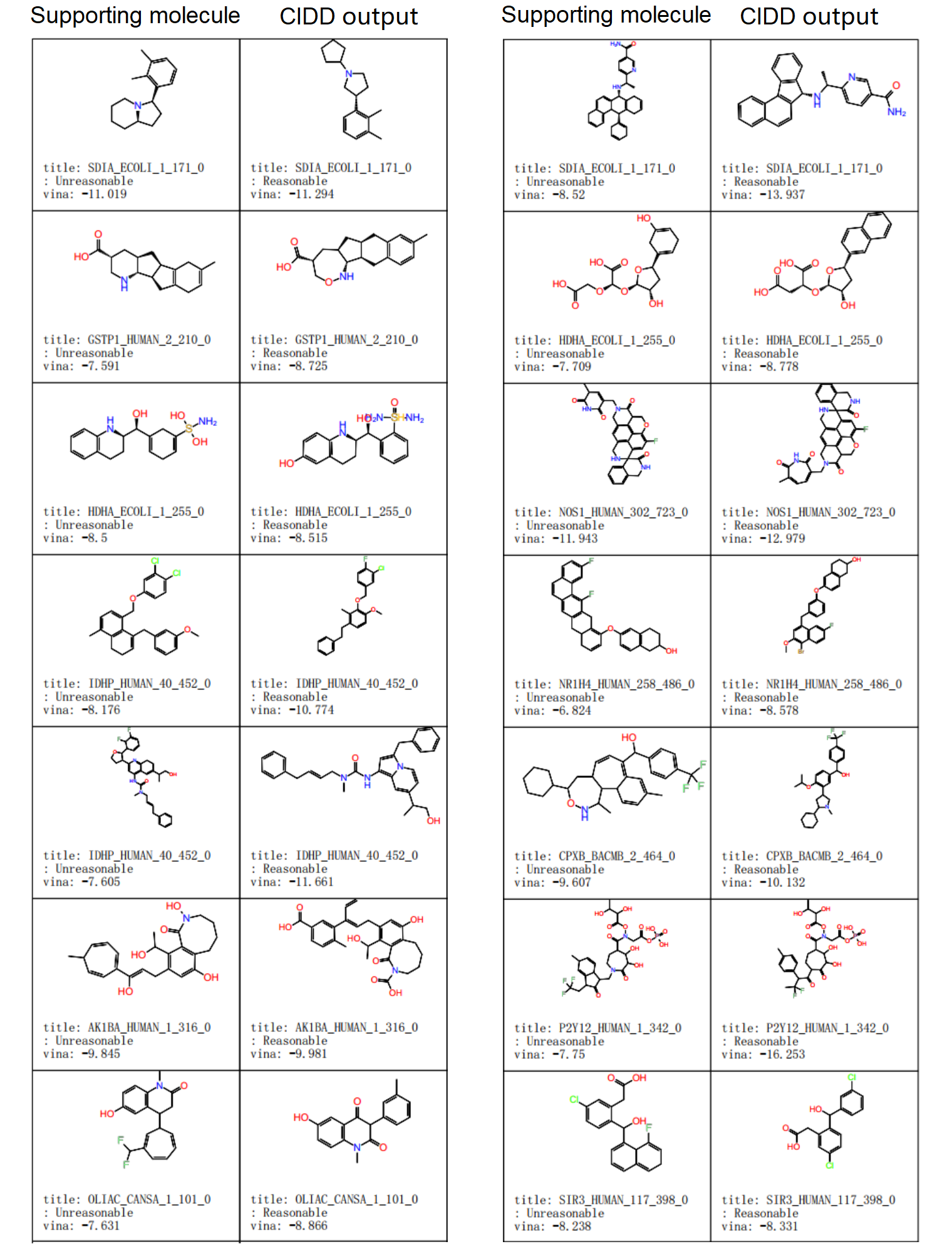}}
% \caption{Interaction analysis module}
\label{fig: case3}
\end{center}
\end{figure}

\begin{figure}[H]
\begin{center}
\centerline{\includegraphics[width=\textwidth]{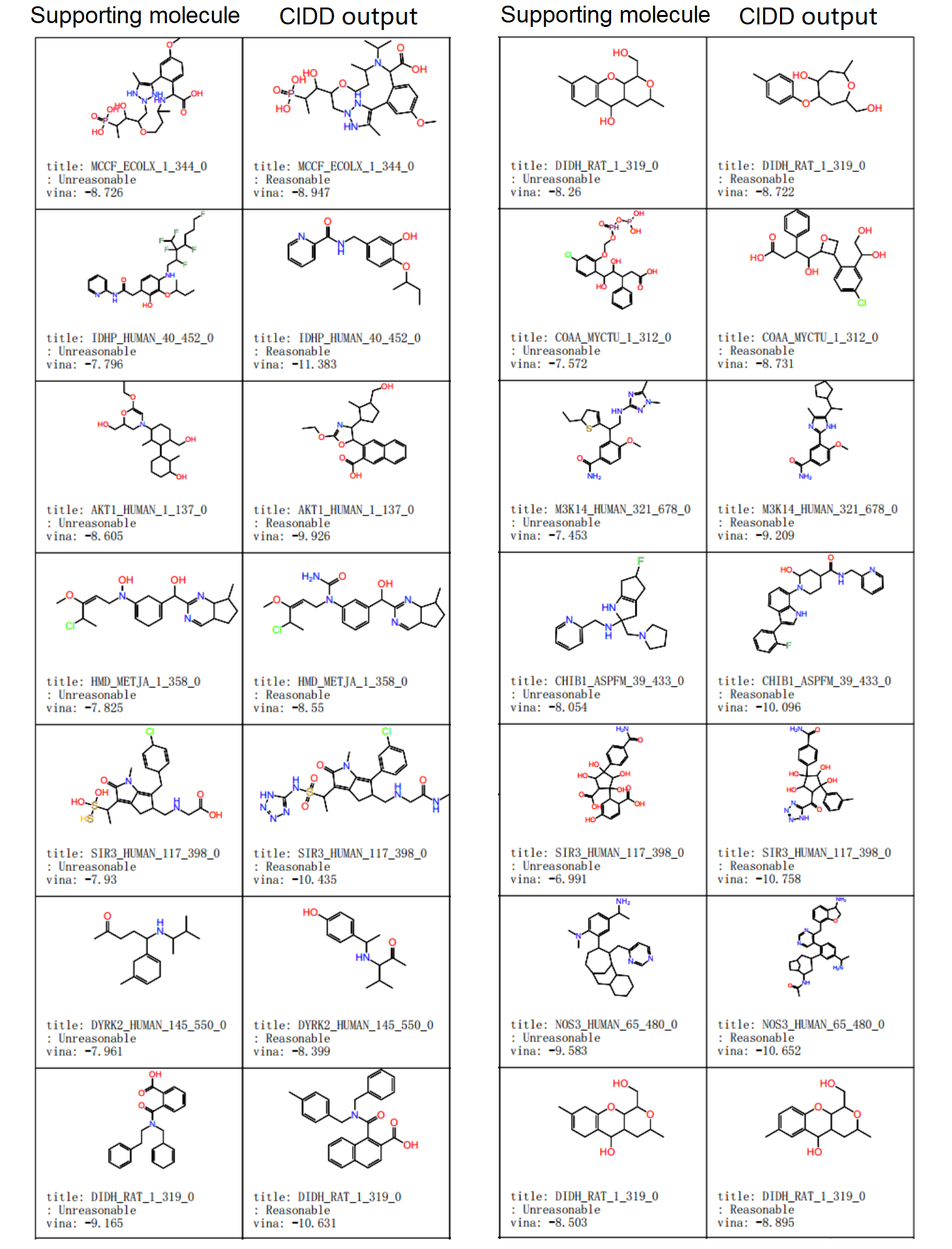}}
% \caption{Interaction analysis module}
\label{fig: case4}
\end{center}
\end{figure}

\begin{figure}[H]
\begin{center}
\centerline{\includegraphics[width=\textwidth]{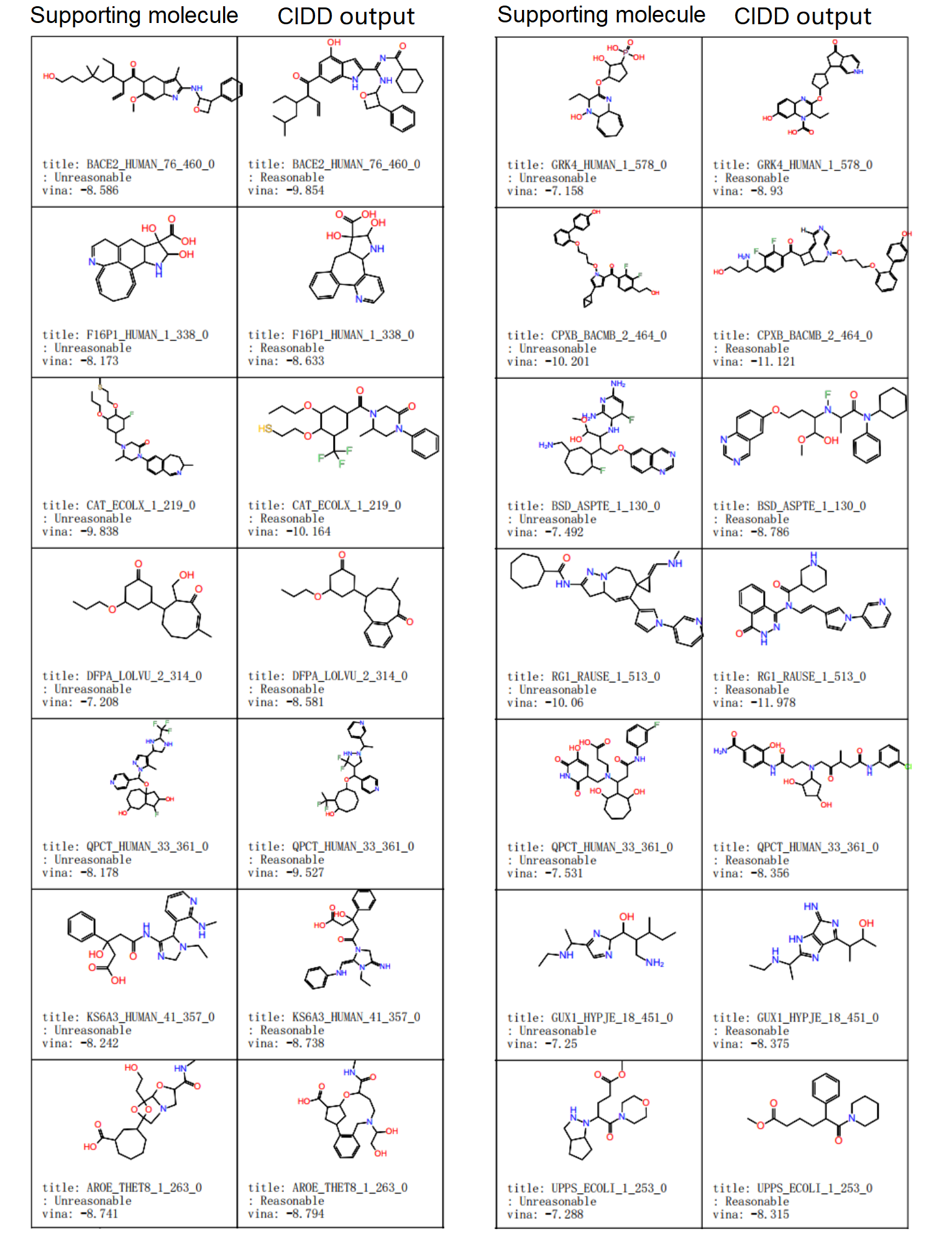}}
% \caption{Interaction analysis module}
\label{fig: case5}
\end{center}
\end{figure}

\begin{figure}[H]
\begin{center}
\centerline{\includegraphics[width=\textwidth]{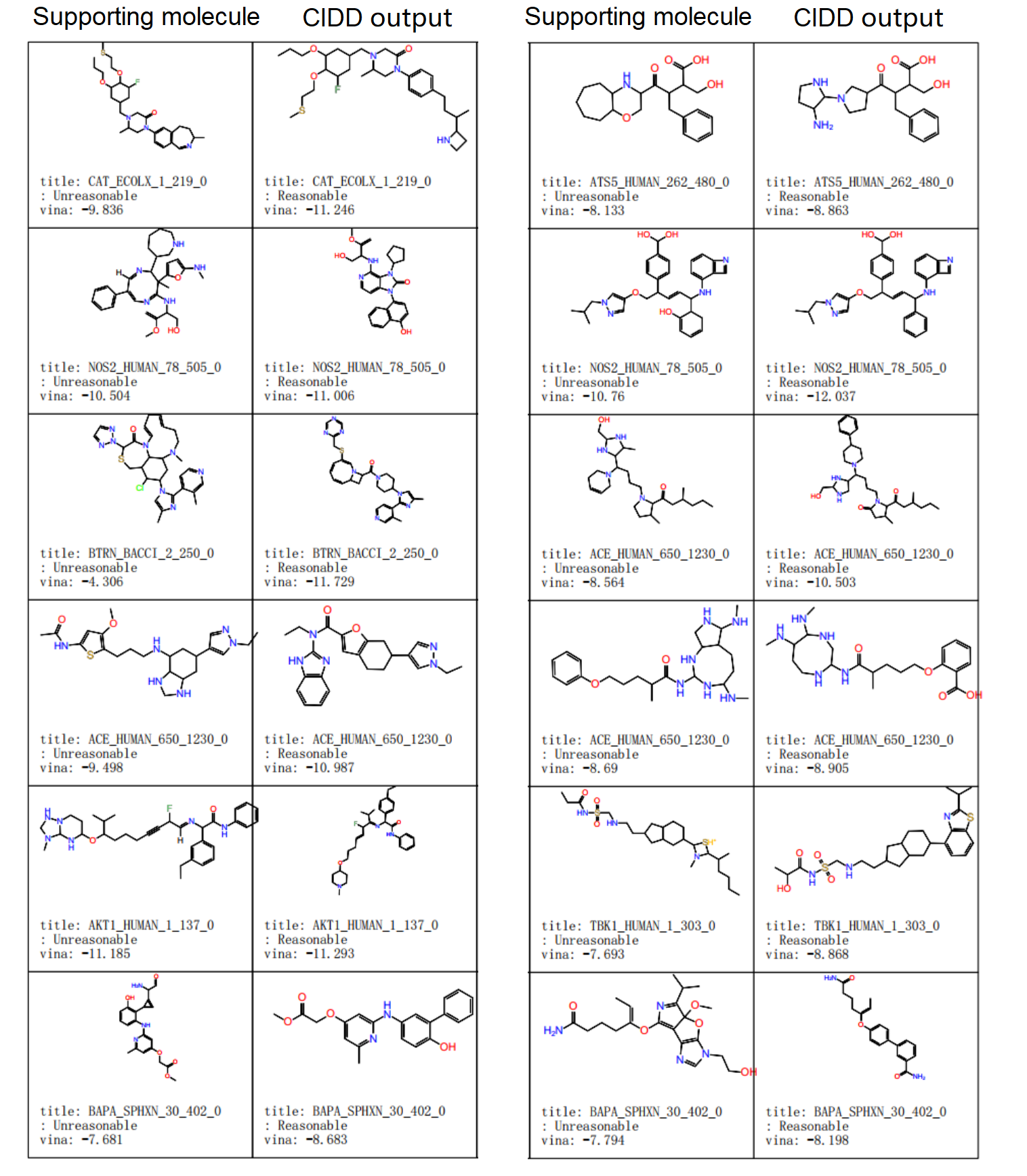}}
% \caption{Interaction analysis module}
\label{fig: case6}
\end{center}
\end{figure}

\begin{figure}[H]
\begin{center}
\centerline{\includegraphics[width=\textwidth]{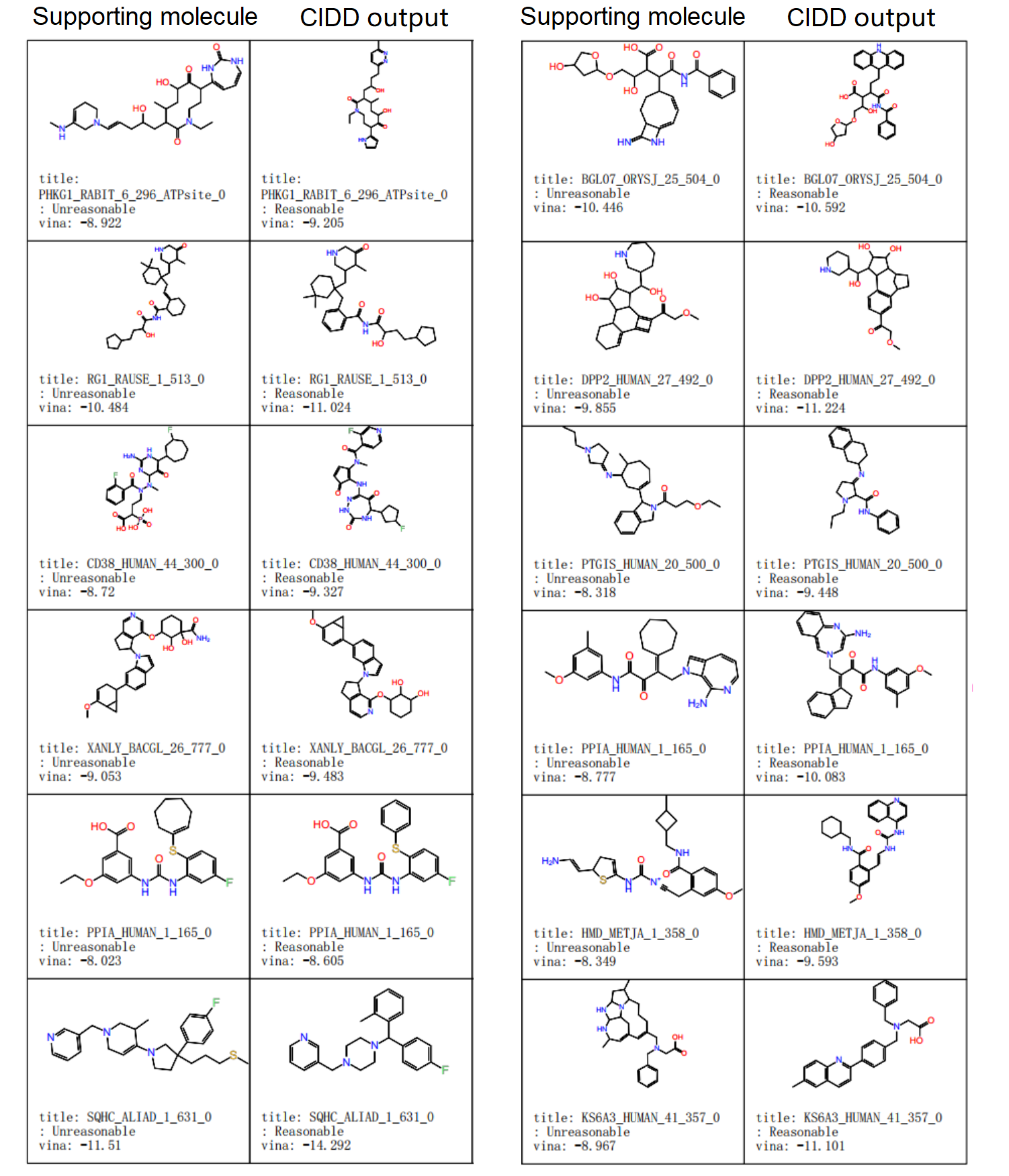}}
% \caption{Interaction analysis module}
\label{fig: case7}
\end{center}
\end{figure}

\begin{figure}[H]
\begin{center}
\centerline{\includegraphics[width=\textwidth]{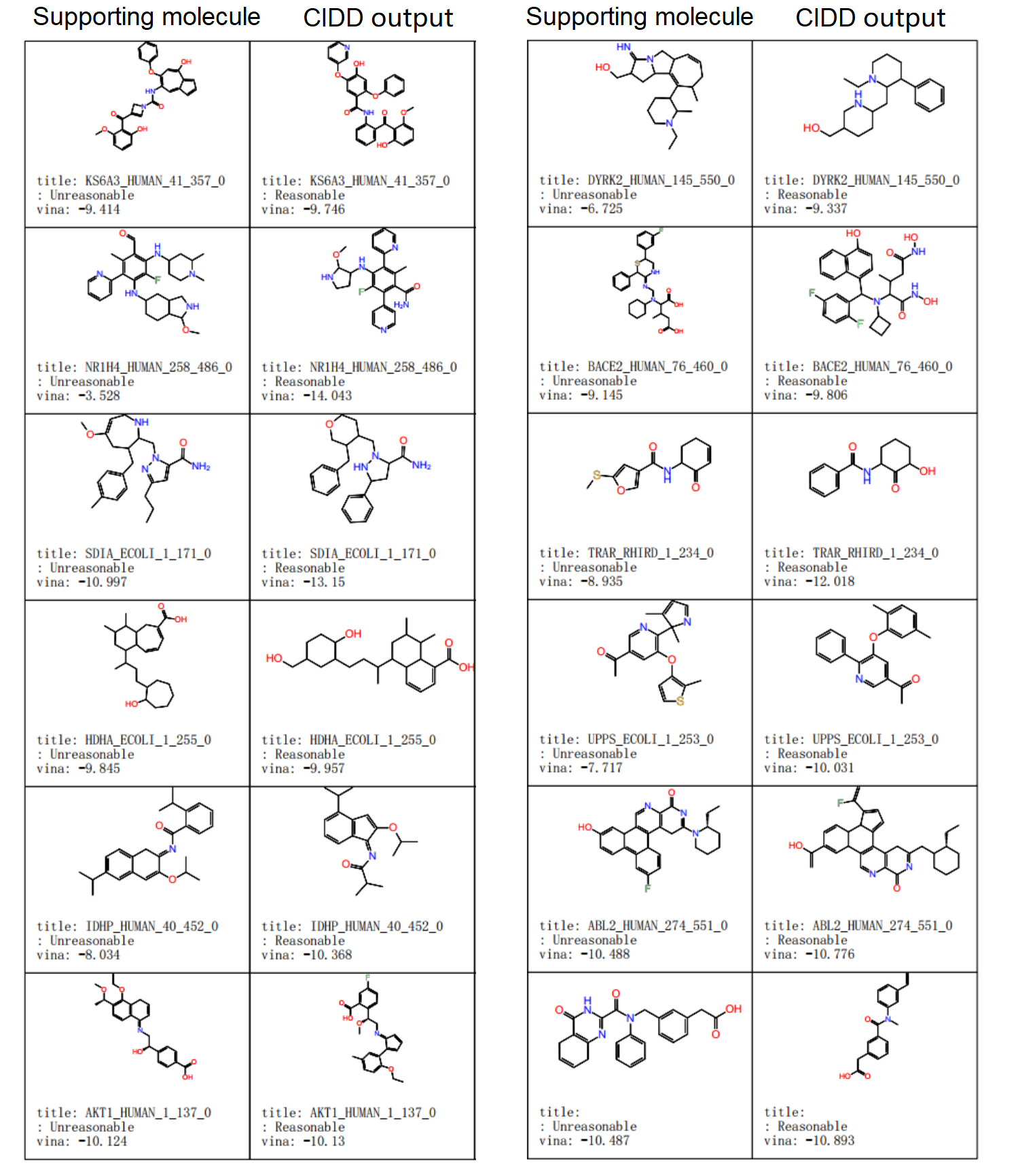}}
% \caption{Interaction analysis module}
\label{fig: case8}
\end{center}
\end{figure}

\begin{figure}[H]
\begin{center}
\centerline{\includegraphics[width=\textwidth]{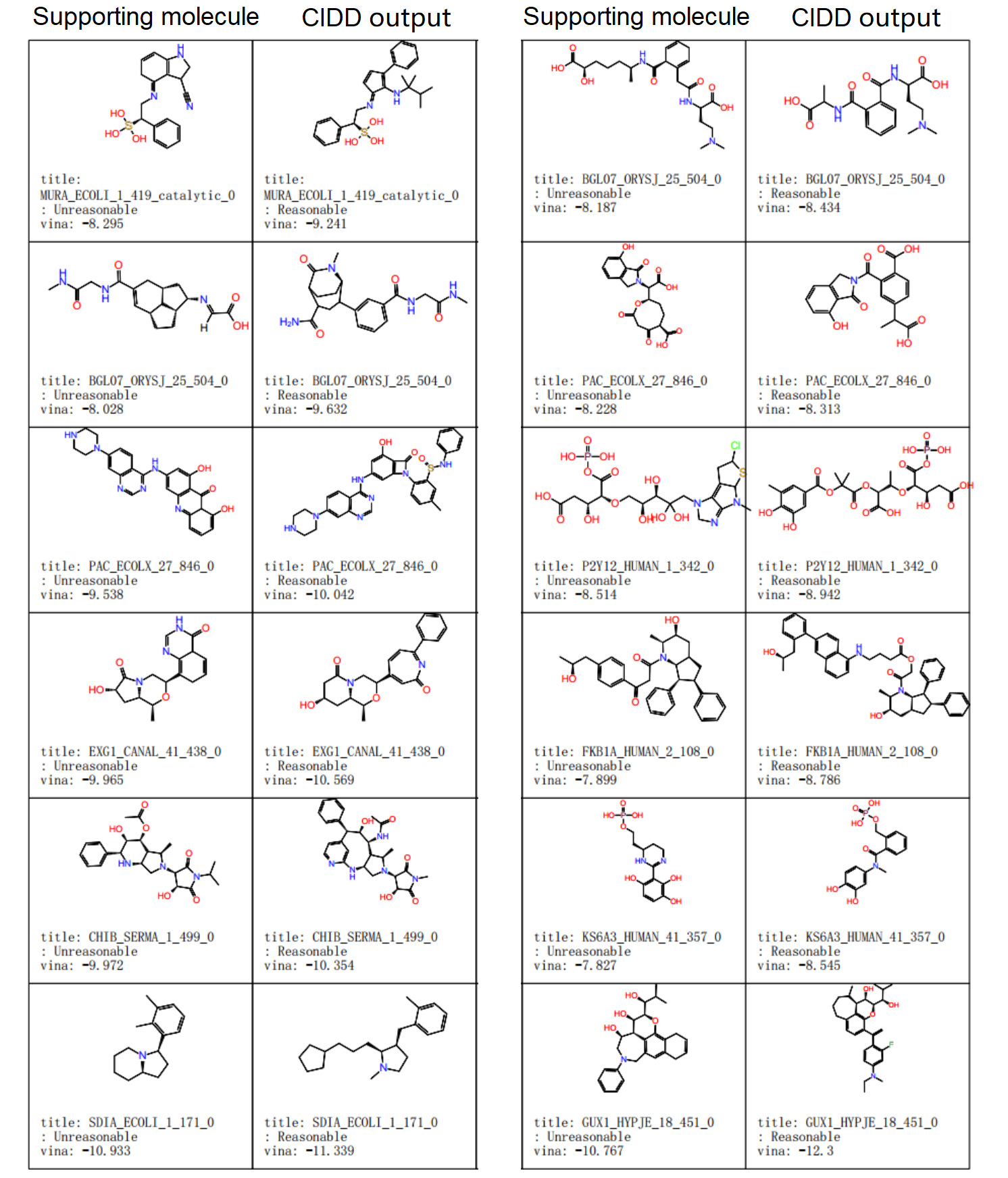}}
% \caption{Interaction analysis module}
\label{fig: case9}
\end{center}
\end{figure}

\begin{figure}[H]
\begin{center}
\centerline{\includegraphics[width=\textwidth]{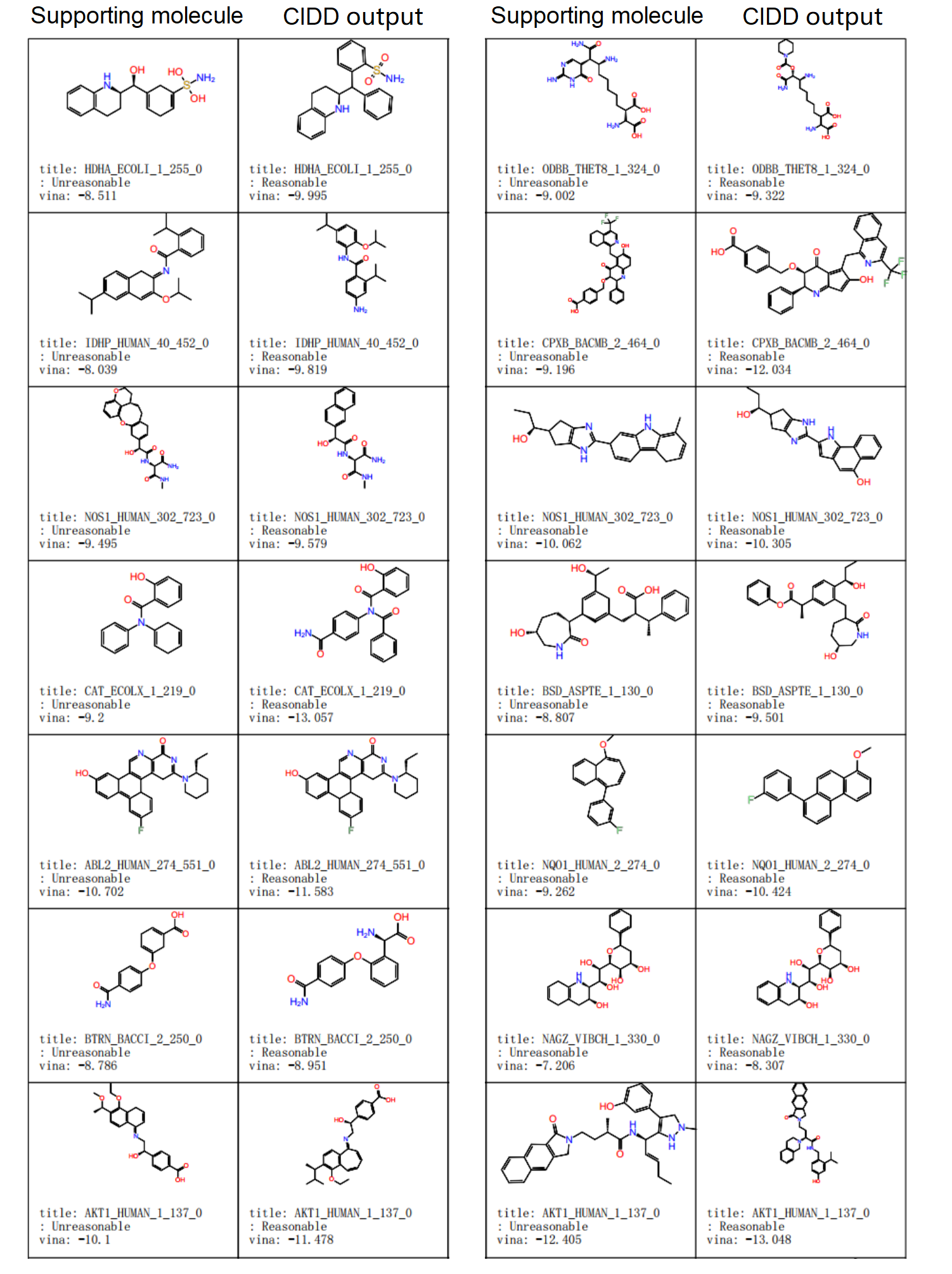}}
% \caption{Interaction analysis module}
\label{fig: case10}
\end{center}
\end{figure}

\begin{figure}[H]
\begin{center}
\centerline{\includegraphics[width=\textwidth]{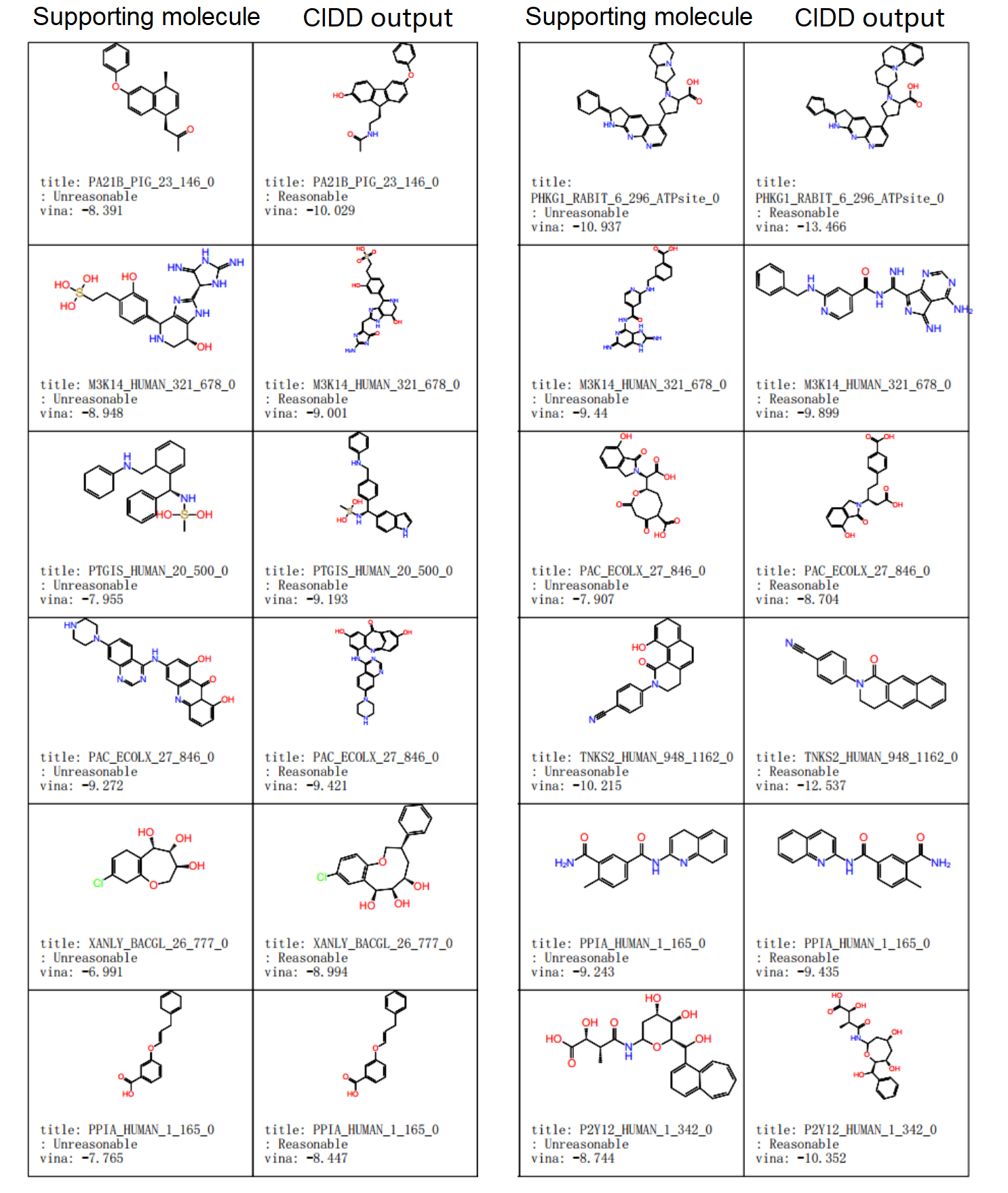}}
% \caption{Interaction analysis module}
\label{fig: case11}
\end{center}
\end{figure}

\begin{figure}[H]
\begin{center}
\centerline{\includegraphics[width=\textwidth]{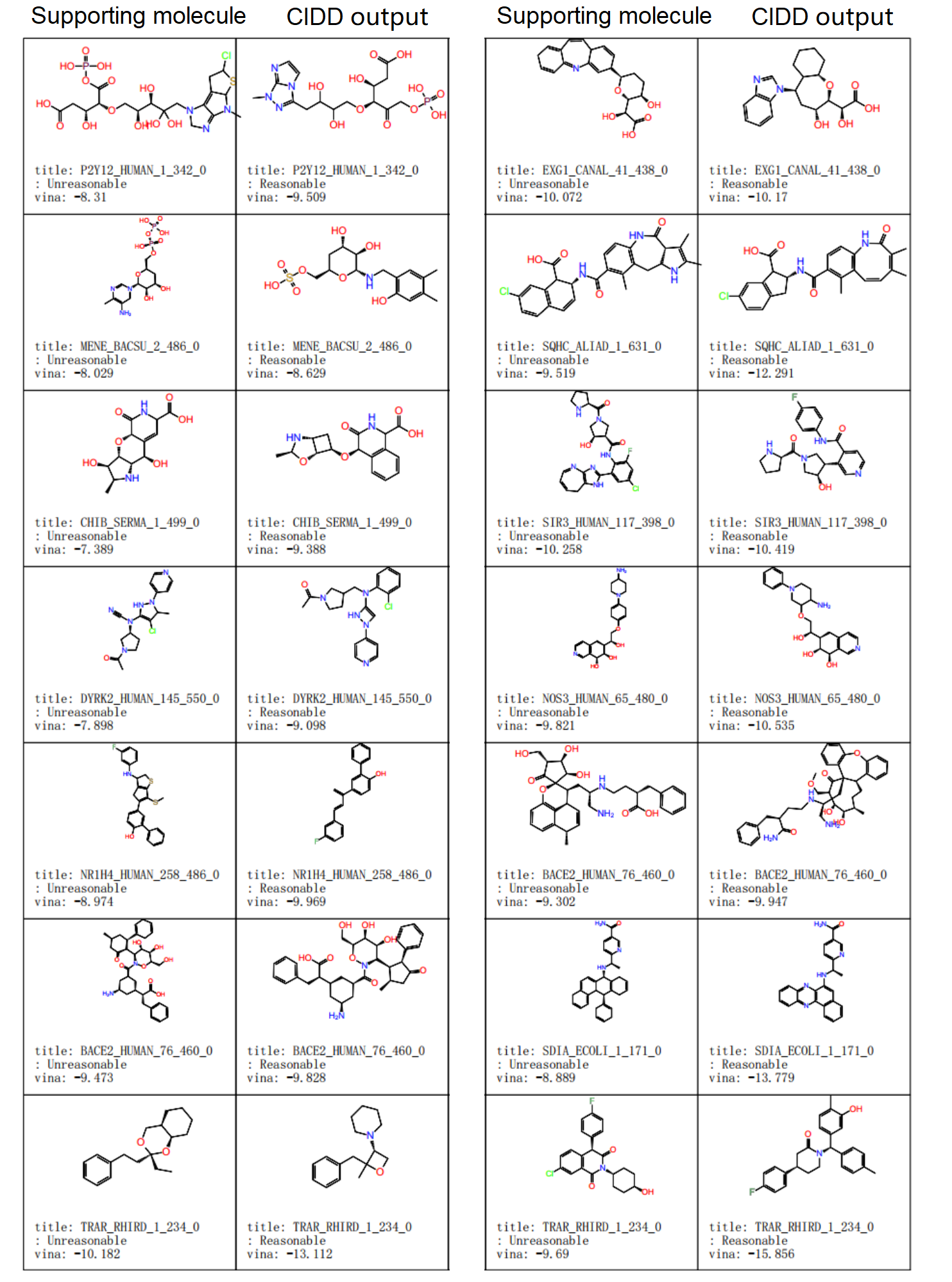}}
% \caption{Interaction analysis module}
\label{fig: case12}
\end{center}
\end{figure}

\begin{figure}[H]
\begin{center}
\centerline{\includegraphics[width=\textwidth]{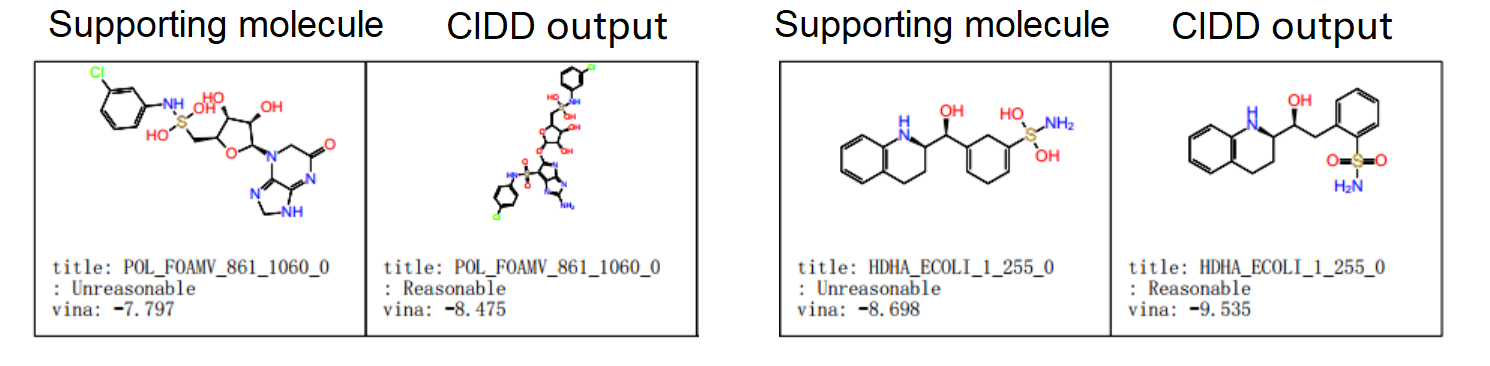}}
% \caption{Interaction analysis module}
\label{fig: case13}
\end{center}
\end{figure}

\end{document}